\documentclass[AMA,Times1COL]{WileyNJDv5} 

\articletype{Article Type}%

\received{Date Month Year}
\revised{Date Month Year}
\accepted{Date Month Year}
\volume{00}
\copyyear{2024}
\startpage{1}

\newcommand{\Y}{{\mbox{\boldmath $Y$}}}

\newcommand{\mbf}[1]{\mbox{\boldmath${#1}$}}
\newcommand{\X}{{\mbox{\boldmath $X$}}}

\newcommand{\Z}{{\mbox{\boldmath $Z$}}}
\newcommand{\U}{{\mbox{\boldmath $U$}}}
\newcommand{\W}{{\mbox{\boldmath $W$}}}
\newcommand{\V}{{\mbox{\boldmath $V$}}}

\newcommand{\bA}{{\mbox{\boldmath $A$}}}

\newcommand{\be}{\mbf{\beta}}

\newcommand{\bgamma}{{\mbox{\boldmath $\gamma$}}}
\newcommand{\bomega}{{\mbox{\boldmath $\omega$}}}

\newcommand{\bSigma}{{\mbox{\boldmath $\Sigma$}}}

\newcommand{\bfeta}{{\mbox{\boldmath $\xi$}}}

\newcommand{\bvarphi}{{\mbox{\boldmath $\varphi$}}}

\hypersetup{
  colorlinks   = true, 
  urlcolor     = black, 
  linkcolor    = black, 
  citecolor   = black 
}

\begin{document}

\title{Quantifying predictive uncertainty of aphasia severity in stroke patients with sparse heteroscedastic Bayesian high-dimensional regression}

\author[1]{Anja Zgodic}

\author[2]{Ray Bai}

\author[1]{Jiajia Zhang}

\author[1]{Yuan Wang}

\author[3]{Chris Rorden}

\author[1]{Alexander C McLain}

\authormark{ZGODIC \textsc{et al}}
\titlemark{Quantifying predictive uncertainty of aphasia severity with H-PROBE}

\address[1]{\orgdiv{Department of Epidemiology and Biostatistics}, \orgname{University of South Carolina}, \orgaddress{\state{South Carolina}, \country{United States}}}

\address[2]{\orgdiv{Department of Statistics}, \orgname{University of South Carolina}, \orgaddress{\state{South Carolina}, \country{United States}}}

\address[3]{\orgdiv{Department of Psychology}, \orgname{University of South Carolina}, \orgaddress{\state{South Carolina}, \country{United States}}}

\corres{Corresponding author Alexander C McLain, 915 Greene Street, Columbia, SC, 29208, USA. \email{mclaina@mailbox.sc.edu}}

\presentaddress{University of South Carolina, 915 Greene Street, Columbia, SC, 29208, USA.}

\abstract[Abstract]{Sparse linear regression methods for high-dimensional data commonly assume that residuals have constant variance, which can be violated in practice. For example, Aphasia Quotient (AQ) is a critical measure of language impairment and informs treatment decisions, but it is challenging to measure in stroke patients. It is of interest to use high-resolution T2 neuroimages of brain damage to predict AQ. However, sparse regression models show marked evidence of heteroscedastic error even after transformations are applied. This violation of the homoscedasticity assumption can lead to bias in estimated coefficients, prediction intervals (PI) with improper length, and increased type I errors. Bayesian heteroscedastic linear regression models relax the homoscedastic error assumption but can enforce restrictive prior assumptions on parameters, and many are computationally infeasible in the high-dimensional setting. This paper proposes estimating high-dimensional heteroscedastic linear regression models using a heteroscedastic partitioned empirical Bayes Expectation Conditional Maximization (H-PROBE) algorithm. H-PROBE is a computationally efficient maximum \textit{a posteriori} estimation approach that requires minimal prior assumptions and can incorporate covariates hypothesized to impact heterogeneity. We apply this method by using high-dimensional neuroimages to predict and provide PIs for AQ that accurately quantify predictive uncertainty. Our analysis demonstrates that H-PROBE can provide narrower PI widths than standard methods without sacrificing coverage. Narrower PIs are clinically important for determining the risk of moderate to severe aphasia. Additionally, through extensive simulation studies, we exhibit that H-PROBE results in superior prediction, variable selection, and predictive inference compared to alternative methods.}


\maketitle

\renewcommand\thefootnote{}
\footnotetext{\textbf{Abbreviations:} AQ, aphasia quotient; PI, prediction interval; TBD, total brain damage; MRI, magnetic resonance imaging; MAP, maximum a posteriori; PX, parameter expanded; EM, expectation-maximimization; ECM, expectation conditional maximization; SNR, signal-to-noise; CV, cross-validation; ECP, empirical coverage probability; RMSE, root mean squared error; MSE, mean squared error; MAD, median absolute deviation; TPR, true positive rate; FDR, false discovery rate; MSPE, mean squared predictive error; LAD, Least Absolute Deviation; LTS, least trimmed squares.}

\renewcommand\thefootnote{\fnsymbol{footnote}}
\setcounter{footnote}{1}

\section{Introduction} \label{sec.intro}

Much of the current literature on sparse linear regression methods for high-dimensional data assumes that the residuals have a constant variance. However, in practice, this assumption is often violated. A clinical application where high-dimensional heteroscedastic data arises is in treatment decisions for neurological disorders based on patient imaging data. Johnson et al\cite{Johnson2019} present results from a study on language rehabilitation in patients who experienced a left-hemispheric stroke and suffer from aphasia -- a language disorder impacting speech. The outcome of interest is the subjects' Aphasia Quotient (AQ), a score quantifying language impairment vital to understanding patients' treatment options.\citep{RisSpr85} However, collecting AQ is a cumbersome task, particularly for patients who have recently had a stroke.\citep{AnsBro05} Consequently, it is of interest to develop models that can \emph{predict} subjects' unknown AQ based on images of their brains.\citep{Lee2021}

While several studies have proposed methods to predict aphasia severity \citep{Yourganov2015, Lee2021, Teghipco2023}, none provide \emph{prediction intervals} (PIs) for AQ predictions. Effective decision-making in healthcare relies crucially on combining predictive models with uncertainty analyses.\citep{Begoli2019, Zou2023} For example, according to the Western Aphasia Battery, the severity of aphasia can be classified as follows: AQ of less than 26 is very severe, 26--50 is severe, 51--75 is moderate, and above 75 is mild.\citep{Kertesz2007} If a patient's PI for AQ spans several categories of aphasia severity, then a clinician could take this predictive uncertainty into account and order additional diagnostic tests or consider other factors in the patient's medical history.\citep{Zou2023} Further, if patients' prediction and PI safely place them in the mild category, their treatment can be streamlined accordingly. Since the PIs can better guide clinicians in defining patient treatment courses, we develop a PI-based approach to quantify predictive uncertainty of AQ. 

\begin{figure}[t]
\centering
\includegraphics[width=5.5in]{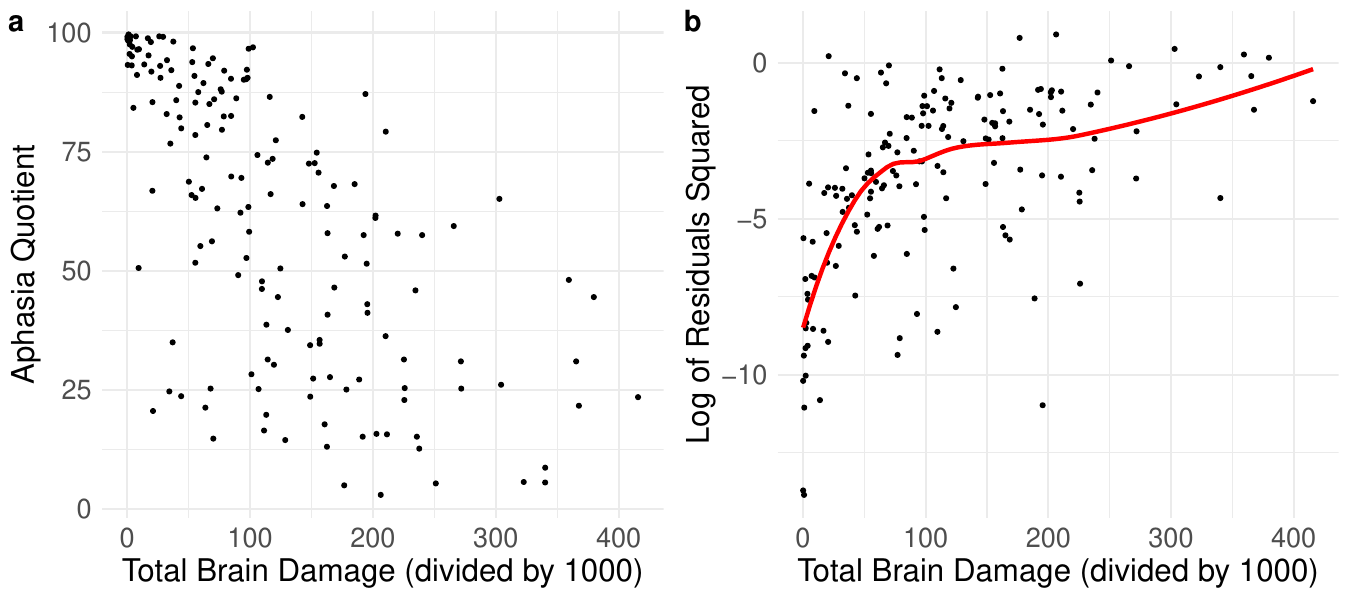} \\
\caption{(a) Aphasia Quotient (AQ) outcome by total brain damage (defined as the number of brain voxels with lesions). (b) Log Residuals Squared from homoscedastic high-dimensional linear regression using brain images as predictors and AQ as the outcome. \label{fig.intro1}}
\end{figure}

\begin{figure}[t]
\centering
\includegraphics[width=5.5in]{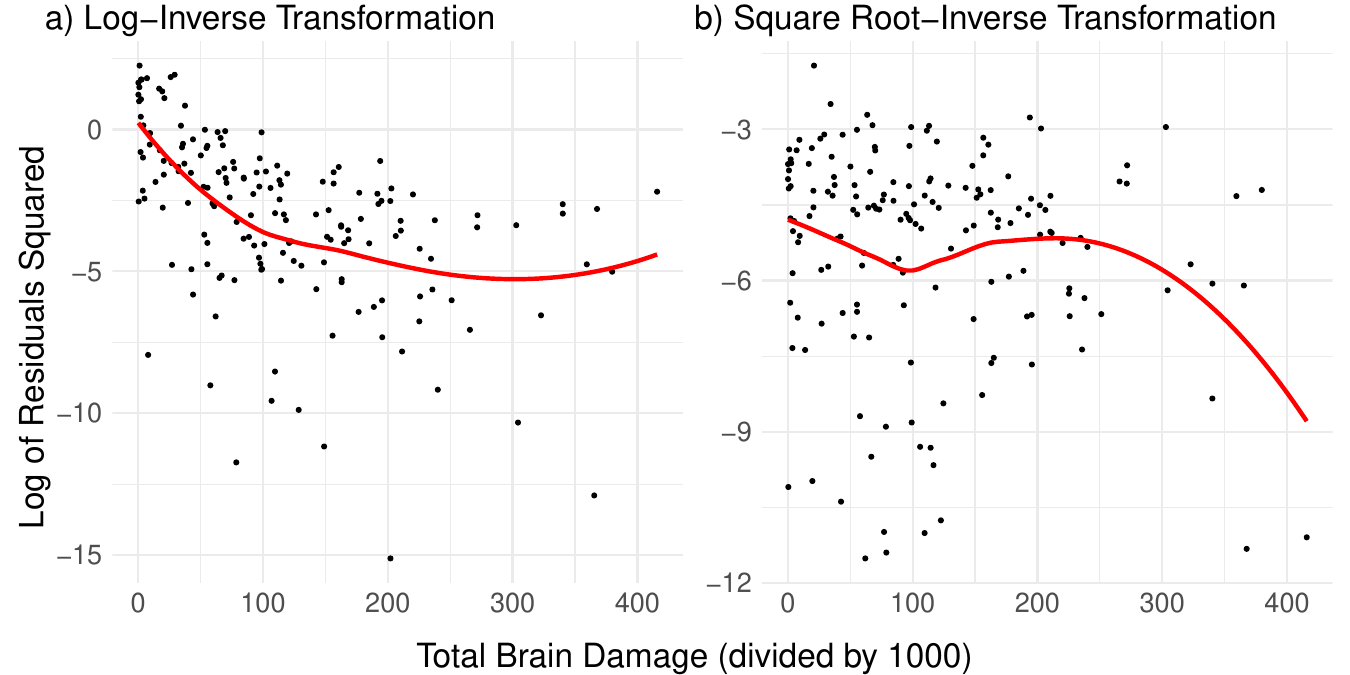} \\
\caption{Log Residuals Squared from homoscedastic high-dimensional linear regression using brain images as predictors with (a) $AQ_{log-inv} = \log AQ_{inv}$, and (b) $AQ_{sqrt-inv} = \sqrt {AQ_{inv}}$ as the outcome where $AQ_{inv}=(100-AQ)/100$.\label{fig.intro2}}
\end{figure}

In our motivating application, brain images are obtained through T1 structural Magnetic Resonance Imaging (MRI) and give the lesion status (i.e., damaged or not damaged from stroke) of more than $5\times 10^6$ three-dimensional ($1\ mm^3$) brain voxels. As displayed in Figure \ref{fig.intro1}a, patients with little brain damage commonly score near the top of the $0-100$ range, while those with more substantial brain damage generally have lower AQ scores. However, there are several examples of patients with small damage and low AQ or large damage and relatively high AQ. Figure \ref{fig.intro1}b suggests that the total number of damaged voxels (total brain damage, TBD) has a positive relationship with the residual variance. A common approach to models exhibiting the aforementioned characteristics (heteroscedasticity and bounded range) is to use a transformed version of the outcome. However, as Figure \ref{fig.intro2} demonstrates, the relationship between model residuals and TBD remains with log and square-root transformations. This finding solidifies the need for a heteroscedastic approach to predict patient AQ scores and provide accurate PIs to help clinical decision-making. 

\subsection{Bayesian high-dimensional heteroscedastic regression} \label{sec.intro.novel}

If ignored, heterogeneity can harm multiple areas of analyses, including bias in estimated coefficients, PIs with improper length, and increased type I errors.\citep{CarRup88} These drawbacks can be particularly impactful in high-dimensional settings where the number of predictors is much larger than the sample size, and heterogeneity can lead to overfitting. Penalized linear regression techniques have been expanded for data displaying non-constant errors. These methods are designed to down-weight outliers or anomalous observations with large error variances. Proposals include methods that weight the objective function of the Least Absolute Shrinkage and Selection Operator (LASSO) \citep{Ziel2016} using Least Absolute Deviation (LAD) \citep{Wang2007} or the square-root of the residual sum of squares \citep{Belloni2014Aim2}, as well as a sparse version of least trimmed squares (LTS).\citep{Rousseeuw2006, Alfons2013}

The aforementioned methods may not be suited for scenarios where known factors are hypothesized to be related to heterogeneity in the data. An instance of these scenarios is our AQ application, where TBD is related to the residual variance. To address this gap, a second line of research specifically focuses on modeling the variance of observations. We have located only four such proposals for the high-dimensional setting, three in the frequentist framework, and one in the Bayesian framework. First, Daye et al\cite{Daye2012} proposed doubly regularized likelihood estimation with $\ell_1$ penalty on parameters for both the mean and variance. Similarly, Chiou et al\cite{Chiou2020} use a greedy algorithm \citep{Temlyakov2000} and backward elimination to select variables for the models on the mean and the variance. Third, Zhou et al\cite{Zhou2021} leverage the  conceptual framework of Daye et al\cite{Daye2012} but use sample splitting to select predictors via LASSO models on the mean and fit additional LASSO models to the residual variance. Finally, Pratola et al\cite{HBART2020} proposed heteroscedastic Bayesian additive regression trees (HBART), where, similarly to Daye et al\cite{Daye2012}, the same predictors are considered for the models on the mean and the variance.

While many of the above methods can sufficiently incorporate heteroscedasticity in the high-dimensional setting, very few of them have investigated creating PIs for future observations. Estimating PIs is a key motivating factor for heteroscedastic regression models since the lengths of PIs may vary by factors available in the data. Conformal inference methods \citep{Vovetal05, Leietal18, TibFoy19} can be used with penalized regression and have finite-sample coverage guarantees. However, such guarantees are marginal and may be higher or lower with some predictor combinations. As demonstrated in our data analysis in Section \ref{sec.da}, marginal properties are unsatisfactory for heterogeneous data, and are particularly inefficient when sources of heterogeneity can be well hypothesized. 

The HBART method can construct PIs for future observations which account for heteroscedasticity.  However, HBART does not perform variable selection and, as illustrated in Section \ref{sec.simul}, is very computationally intensive to fit when the number of predictors $p$ is larger than sample size $n$. As a result, HBART does not result in a parsimonious model. In high dimensions, some form of regularization is often desirable and enhances interpretability. For example, by applying our proposed method to high-dimensional neuroimages, we can identify regions of the brain that are significantly associated with AQ (see Figure \ref{fig.brainmap}).

Given these limitations, we propose a heteroscedastic high-dimensional linear regression model estimated with a Heteroscedastic PaRtitiOned empirical Bayes Expectation conditional maximization (H-PROBE) algorithm. We base H-PROBE on the previously established PROBE framework.\citep{McLain2022} PROBE is a computationally efficient maximum \textit{a posteriori} (MAP) estimation approach based on a quasi Parameter-Expanded Expectation Conditional Maximization (PX-ECM) algorithm.\citep{Meng1993,Liuetal98} It requires minimal prior assumptions on the regression parameters through plug-in empirical Bayes estimates of hyperparameters in the E-step. H-PROBE expands PROBE by allowing for non-constant residual variance via incorporating covariates hypothesized to impact heterogeneity and covariates not subjected to the sparsity assumption. H-PROBE is much less computationally demanding than other Bayesian regression models fitted with Markov chain Monte Carlo (MCMC), such as HBART. Finally, we propose methods to estimate prediction intervals for future observations, which were very infrequently explored in aforementioned works. 

The novel features of H-PROBE are (i) modeling of the mean and variance within the same Bayesian framework, with variable selection in the mean parameters and predictions for new observations, and (ii) addressing a crucial gap in previous heteroscedastic methods, i.e., a lack of procedures to estimate PIs. We demonstrate the utility of incorporating heterogeneity when constructing PIs by analyzing simulated data and our study of AQ in patients with recent left-hemispheric stroke.\citep{Johnson2019} Our work makes both methodological and applied contributions.  On the methodological side, to the best of our knowledge, H-PROBE is the first Bayesian variable selection approach in high-dimensional modeling which includes both models on the mean and the variance. On the application side, we expand the literature on AQ prediction by outlining the first proposal that provides PIs and AQ predictions. The remaining paper layout is as follows. We describe H-PROBE in Section \ref{sec.methods} and numerical studies evaluating its performance in Section \ref{sec.simul}. Section \ref{sec.da} presents a comprehensive analysis of the AQ imaging study, and Section \ref{sec.discuss} concludes with a brief discussion. Supplementary Materials include more detailed technical content, additional simulations, and further data analysis results.

\section{Methods}\label{sec.methods}

\subsection{Model framework}\label{sec.blmm.setup}

In most linear regression models with outcome $\Y = (Y_1 ,Y_2, \ldots, Y_n)$, predictors $\X = (\X_1, \ldots, \X_p)$, and error term $\mbf{\epsilon} = (\epsilon_1, \epsilon_2, \ldots, \epsilon_n)$, it is assumed that $Var(\epsilon_i) = \sigma^2$ for all $i=1, \ldots, n$. As a result, the error term has the same variance for all observations (homoscedasticity). In a contrasting scenario, the error term may display a variance that differs from observation to observation. Then, the linear regression model for heteroscedastic data is written as
\begin{equation}\label{eq.lm}
Y_{i} = \X_{i} \mbf{b}+ \epsilon_{i},
\end{equation}
where $\mbf{b} \in \mathcal{R}^{p}$, $E(\epsilon_i) = 0$, and $Var(\epsilon_{i}) = \sigma^2_i$. Let $X_{ik}$ represent predictor $k$ for observation $i$, with $n \times 1$ vector $\X_k = (X_{1k}, \ldots, X_{nk})$, $n \times p$ design matrix $\X$, and $Var(\mbf{\epsilon}) = \bSigma = diag(\sigma^2_1, \ldots, \sigma^2_n)$ as a diagonal $n \times n$ matrix. Assuming Gaussian errors the distribution of the outcome is $\Y \sim N(\X \mbf{b}, \bSigma)$.  

We leverage a Bayesian framework to accommodate a high-dimensional ($p \gg n$) setting and conduct sparse linear regression. Specifically, our model will allow for sparse and non-sparse predictors. For ease of presentation, we assume the mean's non-sparse predictors are also the variance predictors, denoted by $\V \in \mathcal{R}^{n \times v}$. In practice, this need not be the case.  With this we rewrite the model in (\ref{eq.lm}) as 
\begin{equation}\label{eq.mod}
\Y = \X (\bgamma \circ \be) + \V \bvarphi + \mbf{\epsilon},
\end{equation}
where $\bgamma \circ \be$ is a Hadamard product, $\bgamma \in \{0,1\}^p$, and $\bvarphi \in \mathcal{R}^{v}$. For brevity, let $\bgamma  \be \equiv \bgamma \circ \be$ for the remainder. Let $\mbf{\mathcal{D}} = \{\mathcal{D}_1,\ldots,\mathcal{D}_n\}$ with $\mathcal{D}_i = (Y_i,\X_i,\V_i)$ denote the observed data. We add the following parametric model on the diagonal variance matrix $\bSigma$,
\begin{eqnarray}\label{eq.modvar}
-\log\{diag(\bSigma)\} &=& \V\bomega, 
\end{eqnarray}
where  $\bomega \in \mathcal{R}^{v}$. The log transformation on the variances ensures positivity, can accommodate variances that vary over orders of magnitude, and has been long established in variance function modeling.\citep{Carroll1988, Cleveland1993} The complete Bayesian framework includes prior information on
\begin{eqnarray*}
p(\be ) &=& \prod_{k=1}^p f_\beta(\beta_k), \\
p(\bgamma|\pi) &=& \pi^{p-|\gamma| } (1-\pi)^{|\gamma|}, \\
p(\bvarphi ) &\propto& 1, \\
\bomega &\sim& \mbox{MLG}(\mbf{0}, c^{1/2} \sigma^2_\omega \textit{\textbf{I}}, c\mbf{1}, c\mbf{1}), 
\end{eqnarray*}
where MLG denotes a Multivariate Log-Gamma distribution with $c,\sigma^2_\omega>0$.\citep{Parker2021} Throughout, we use $\sigma^{-1}_{\omega} = 0$ and $c = 1000$ to yield a weakly informative prior on $\bomega$.

We use an ECM algorithm to estimate parameters of interest in the model framework presented above. Within the ECM algorithm, the parameter estimates at iteration $t$ are used to obtain the expected complete-data log-posterior distribution with respect to $\bgamma$,
\begin{equation}\nonumber
E_{\bgamma}\left\{ \log \ p( \be, \bvarphi, \bomega | \mbf{\mathcal{D}}, \bgamma)| \mbf{\mathcal{D}}, \be^{(t)}, \bvarphi^{(t)}, \bomega^{(t)} \right\}.
\end{equation}
In the algorithm, the CM-steps consist of calculating the MAP values for each parameter conditional on the previous values of the remaining parameters. This amounts to calculating the modes of the expected log-conditional posterior distributions, which we derive below. 

To obtain conditional posterior distributions for all parameters, we assume $f_{\beta} \propto 1$ and $\pi \sim \mbox{Uniform}(0,1)$. In practice, we leave hyperparameters $\pi$ and $f_{\beta}$ unspecified and estimate them using plug-in empirical Bayes estimators (details in Supplementary Materials). To show the conditional posterior distribution of $\be$, we split $(\X, \be)$ into $(\X_{\gamma}, \be_{\gamma})$ when $\gamma_k = 1$ and $(\X_{\bar{\gamma}}, \be_{\bar{\gamma}})$ when $\gamma_k = 0$.  Conditional on $\bgamma$ and $\bomega$ (i.e., $\bSigma$), the posterior distribution of $\bgamma \be$ and $\bvarphi$ is 
\begin{equation*}\label{eq.post.beast}
\begin{pmatrix}
    \be_\gamma \\ \bvarphi
\end{pmatrix}
 \bigg|(\mbf{\mathcal{D}},\bomega,\bgamma) \sim N\left\{   (\U_\gamma' \bSigma^{-1} \U_\gamma)^{-1} \U_\gamma' \bSigma^{-1} \Y ,(\U_\gamma' \bSigma^{-1} \U_\gamma)^{-1} \right\}
\end{equation*}
where $\U_\gamma = (\X_{\gamma} \ \V)$, while $\be_{\bar{\gamma}}|(\mbf{\mathcal{D}},\bomega,\bgamma) \sim \delta_0(\cdot)$ a point mass at zero. Given $\be$, $\bgamma$, and $\bvarphi$ the posterior for $\bomega$ has density 
\begin{equation}
    f_{\omega}(\bomega) \propto \exp \left\{ \mbf{c}'_{\omega} \mbf{H}_{\omega} \bomega - \mbf{\kappa}'_{\omega} \exp\left( \mbf{H}_{\omega} \bomega \right) \right\}, \label{eq.post.mlg}
\end{equation}
which is proportional to an $MLG(\mbf{0}, \mbf{H}_{\omega}, \mbf{c}_{\omega}, \mbf{\kappa}_{\omega})$ distribution with 
\begin{gather}
\mbf{H}_{\omega} = 
\begin{bmatrix}
\V \\
c^{-1/2} \sigma^{-1}_{\omega} \mbf{I}_{v} 
\end{bmatrix}, 
\ \mbf{c}_{\omega} = \left( \frac{1}{2} \mbf{1}_n', c \mbf{1}_{v}' \right)', \ \mbox{and} \  
\mbf{\kappa}_{\omega} = \left( \frac{1}{2} ||\Y - \X (\bgamma \be) - \V \bvarphi||^2, c \mbf{1}_{v}' \right)', \nonumber 
\end{gather}
where $\mbf{I}_{v}$ denotes a $v \times v$ identity matrix, $\mbf{1}_{v}$ a $v \times 1$ vector of ones, and $||\mbf{x}||^2=\mbf{x}\mbf{x}$ is a Hadamard product. Finally, for $\bgamma$, the E-step results in estimates of the posterior expectation $\mbf{p} = (p_1, \ldots, p_p)$ where $p_k = P(\gamma_k = 1 | \mbf{\mathcal{D}}, \pi)$. Derivations and complete estimation details are provided in Supplementary Materials, while an overview is provided below.

\subsection{Estimation overview}\label{sec.part.setup}

We begin this Section with an overview of the PROBE algorithm, leaving much of the statistical motivations to McLain et al.\cite{McLain2022} The aim of PROBE is to perform MAP estimation of $\bgamma$, $\be$, and $\bvarphi$ in a high-dimensional setting. PROBE is based on a quasi Parameter-Expanded Expectation-Conditional-Maximization (PX-ECM) algorithm, which is a combination of the ECM  and PX-EM algorithms.\citep{Meng1993,Liuetal98} The CM-step of the PX-ECM results in coordinate-wise (predictor-specific) optimization, where the remaining parameters are restricted to their values at the previous iteration. As a result, we partition the mean of the model (\ref{eq.mod}) by predictor $k$. 

The notation $\bA_{\char`\\ k}$ indicates the matrix, vector, or collection without predictor or element $k$. We define $\W_k = \X_{\char`\\ k}(\bgamma_{\char`\\ k} \be_{\char`\\ k}) + \V\bvarphi = (W_{1k}, \ldots, W_{nk})'$ which gives 
%
$E(\Y|\W_k) =   \X_k \beta_k + \W_k$, and
%
$\W_k$ encompasses the impact of all predictors except $\X_k$. 
Since $\W_k$ is estimated, we use parameter-expansion \citep{Liuetal98} to include coefficient $\alpha_k$ which adjusts for the impact of $\W_k$ when updating $\beta_k$ for all $k=1,\ldots,p$. This results in the model
\begin{eqnarray}\label{eq.expand.part}
E(\Y|\W_k) =   \X_k \beta_k + \W_k \alpha_k.
\end{eqnarray}
In H-PROBE, the estimator
to obtain the MAP of $\beta_k|(\gamma_k=1)$ and $\alpha_k$ is $(\hat{{\beta}}_k, \hat{\alpha}_k)'= (\Z_k' \bSigma^{-1} \Z_k)^{-1}\Z_k' \bSigma^{-1} \Y$, where $\Z_k = (\X_k \ \W_k)$. 
In the CM-step for $\beta_k$, we condition on $(\be_{\char`\\ k},\bvarphi)$. However, $\W_k$ is still unknown since it is a function of ``missing data'' $\bgamma_{\char`\\ k}$ and the expectation of $\W_k$ and $\W^2_k$ are required to proceed.

The aim of the E-step is to obtain updates for $E(\W_k)$ and $E(\W^2_k)$ for all $k$, where the expectations are over $\bgamma_{\char`\\ k}$. This requires estimates $p_k$ for all $k$, which are made via plug-in empirical Bayes estimators of $\pi$ and $f_{\beta}$. This procedure requires estimates of the posterior covariance of $\phi$ and the posterior variance $\beta_k|\gamma_k=1$ for all $k$, which are obtained by assuming their marginal posteriors are Gaussian. The $p_k$ estimates update the moments $\W_k$ for all $k$. Further, we estimate the moments of the overall effect of $\X$, denoted by $\W_0 = \X \bgamma \be$. 

The partitioned and overall expectations are then used to perform the subsequent CM-steps, where MAP estimates are updated for $\be$, $\bvarphi$, and $\bomega$ with the new expectations. The MAP of $\bomega$ amounts to maximizing its expected log-conditional posterior density, i.e., the log of (\ref{eq.post.mlg}), which does not have a closed form. As a result, the MAP of $\bomega$ is obtained via quasi-Newton optimization.\citep{Fletcher1987} More discussion contrasting PROBE to other extensions of the EM algorithm is available in McLain et al.\cite{McLain2022} All technical details to perform the estimation procedures are provided in Supplementary Materials.

\subsection{Estimates and model checks}\label{sec.conv.pred}

The H-PROBE method converges when subsequent changes in the expected $\W_0$ values are small, as they capture changes in all regression parameters. 
Upon convergence, H-PROBE provides the MAP and empirical Bayes estimates for parameters described in Section \ref{sec.part.setup}. Table \ref{methods.table.est} provides a summary of the critical parameters, their estimates, as well as their definitions. 
In Numerical Studies (Section \ref{sec.simul}) and Data Analysis (Section \ref{sec.da}), we use $ \tilde{\alpha} (\tilde{\mbf{p}} \tilde{\be})$ -- a combination of $E(\bgamma \be)$ and the MAP of $\alpha$ -- to estimate the impact of sparse predictors $\bgamma \be$. While the properties of non-sparse predictor coefficients $\tilde{\bvarphi}$ are not the focus of this research, we do wish to account for the uncertainty they contribute to the PIs constructed below. We use $\tilde \Psi$ to designate the estimated posterior covariance of $(\tilde{\bvarphi},\tilde{\alpha})$, which we use in formulating PIs in Section \ref{sec.pred.int}.

\renewcommand{\arraystretch}{1.3}
\begin{table}[t]
\caption{Parameters and estimates provided by H-PROBE. SM = Supplementary Materials.}
\label{methods.table.est}
\begin{center}
\begin{tabular}{||p{1.3cm} p{1.2cm} p{1.2cm} p{8cm} ||} 
 \hline
 Parameter & Estimate & Equation & Parameter Definition \\ 
 \hline\hline 
 \space $\be$ & $\tilde{\be}$ & (\ref{eq.mod}), (\ref{eq.expand.part}) & Vector of sparse regression coefficients for predictors in $\X$, conditional on $\bgamma = 1$, in the model on the mean. \\
 \hline
 $\mbf{S}^2$ & $\tilde{\mbf{S}}^2$ & SM & Vector of posterior variances of $\be|(\bgamma = 1)$. \\ 
 \hline
 $\bgamma$ & $\tilde{\mbf{p}}$ & (\ref{eq.mod}), (\ref{eq.expand.part}) & Vector of inclusion indicators for sparse coefficients $\be$ associated with predictors in $\X$. \\
 \hline
 $\bvarphi$ & $\tilde{\bvarphi}$ & (\ref{eq.mod}) & Vector of non-sparse regression coefficients in the model on the mean. \\
 \hline
 $\W_0$ & $\tilde{\W}_0$ & (\ref{eq.expand.part}) & Overall (non-partitioned) latent parameter $\W_0 = \X(\bgamma \be)$. \\
 \hline
 $\alpha$ & $\tilde{\alpha}$ & (\ref{eq.expand.part}) & Overall (non-partitioned) coefficient adjusting for the impact of overall (non-partitioned) $\W_0$. \\
 \hline
 $\bomega$ & $\tilde{\bomega}$ & (\ref{eq.modvar}) & Vector of non-sparse regression coefficients in the model on the variance. \\ 
 \hline
 $\bSigma$ & $\tilde{\bSigma}$ & (\ref{eq.modvar}) & Diagonal variance matrix $\bSigma = \exp\{-(\V \bomega)\}$. \\ 
 \hline
\end{tabular}
\end{center}
\end{table}

The residuals from a high-dimensional homoscedastic regression model can be used to identify variables that may have an association with the residual variance and appropriate forms of such variables such that the parametric assumptions of the variance model in (\ref{eq.modvar}) are adequately met. Supplementary Materials provide additional information and examples for model checks.

\subsection{Prediction intervals}\label{sec.pred.int}

This research aims to develop point estimates and PIs for a future observation not included in the training set with predictor data $\X_{new}$ and $\V_{new}$. MAP estimation does not provide posterior distributions of model parameters, just their mode, limiting predictive inference capabilities. As a result, we assume that the estimates of the posterior variance of $\tilde{\bvarphi}$ and $\tilde{\beta}_k|\gamma_k=1$ for all $k$ can be used to capture the posterior variability of these parameters. To predict for a new observation we use the MAP of the predicted value $ \tilde{Y}_{new} = \V_{new}\tilde\bvarphi + \tilde{W}_{0new}\tilde{\alpha}$, where $\tilde{W}_{0new} =  \X_{new} (\tilde{\mbf{p}} \tilde{\be})$. Further, the variance of $\tilde{W}_{0new}$ is estimated with $\tilde{\mathcal{V}}_{new}  = \X_{new}^2\left\{\tilde{\mbf{p}}\tilde{\mbf{S}}^2 +  \tilde{\be}^{2}  \tilde{\mbf{p}}(1- \tilde{\mbf{p}}) \right\}$. To estimate the variance of $\tilde{Y}_{new}$ while acknowledging the uncertainty in $\tilde{W}_{0new}$, we use 
\begin{eqnarray*}
Var(\tilde{Y}_{new}) = \tilde{\Z}_{new}' \tilde \Psi \tilde{\Z}_{new} + \tilde{\mathcal{V}}_{new}\left\{Var(\tilde{\alpha}) + \tilde{\alpha}^{2}\right\},
\end{eqnarray*}
where $\tilde{\Z}_{new} = (\V_{new}, \tilde{W}_{0new})'$, which is motivated by the measurement error literature.\citep{Buo95} Prediction intervals (PIs) can be formed using the appropriate critical values with $Var(\tilde{Y}_{new})+\tilde{\sigma}^2_{new}$ where $\tilde{\sigma}^2_{new} = \exp\{-(\V_{new}' \tilde{\bomega})\}$ denotes the MAP estimate of the variance for a \textit{new} subject. In the following sections, we evaluate the empirical coverage probabilities of PIs using this approach for test data via simulation studies and a clinical application.

\section{Numerical Studies}\label{sec.simul}

We perform numerical studies to evaluate the performance of H-PROBE. We generate the outcome using $Y_i = \X_i' (\bgamma \be) + \epsilon_i$, where $\epsilon_i \sim N(0, \sigma^2_i)$, $\sigma^2_i = \exp(-\V_i' \bomega)$, $\be \sim U(0, 2\eta_{\beta})$, and $\omega_j=\bar \omega$ for all $j$. We set $\bar \omega$ such that the expected signal-to-noise ratio ($SNR$) is $SNR = 1$ or $2$, where $SNR = E\{Var(\X_i'\bgamma \be)/\exp(-\V_i' \bomega)\}$. We generated correlated continuous predictors $\X_i \sim MVN(\mbf{0}+a_{i},\Sigma)$ where $a_{i} \sim N\left(0, \frac{3}{4}\right)$ and $\Sigma$ is a squared exponential covariance function. Specifically, all predictors are superimposed on a $\sqrt{p} \times \sqrt{p}$ grid, where $\mbf{d}_k = (d_{1k}, d_{2k})$ denotes coordinates of $X_k$. The $(k,k')$ element of the covariance matrix is $\exp\{- \vert\vert (\mbf{d}_k - \mbf{d}_{k'})/20 \vert\vert_2^2 \}$ where $\vert\vert \cdot \vert\vert_2$ denotes the $\ell_2$-norm. 

Our correlation-inducing concept is motivated by Gaussian random fields \citep{Schlather2015}, and mimics the real-world AQ application, where brain voxels closer together are more correlated than voxels further away from each other. Similarly, the $SNR$ settings mirror the AQ application, and in some scenarios make estimation more challenging (i.e., those with lower $SNR$). We also tested correlated binary predictors, generated by applying the indicator that the continuous predictors are less than zero. $\bgamma$ was generated similarly to the binary predictors variables, such that $\sum \gamma_k = p\pi$ in each iteration. Finally, we generated $\V$ such that it included an intercept along with an equal number of standard normal and Bernoulli$(0.5)$ predictor variables.

Simulation settings were varied by the number of predictors in $\X$, $p = (20^2, 75^2)$, the number of predictors in $\V$ including an intercept, $v = (3, 7)$, the proportion of non-zero coefficients, $\pi = (0.01, 0.05)$, the signal-to-noise ratio, $SNR = (1, 2)$, and the average effect size of $\be$, $\eta_{\beta} = (0.3, 0.8)$. For brevity, we focus below on the results for $\eta_{\beta} = 0.8$ and binary $\X$ predictors; results for $\eta_{\beta} = 0.3$ and continuous $\X$ were overwhelmingly similar and are omitted. All simulations had $N=400$ observations and were repeated $400$ iterations. We compare H-PROBE to PROBE and LASSO for all settings. We also compare H-PROBE to heteroscedastic BART (HBART), a Bayesian linear model with a horseshoe prior \citep{CarvahoPolsonScott2010}, and an empirical Bayes approach for prediction in sparse high-dimensional linear regression (EBREG) \cite{MarTan20}. Due to the high computational cost, we only ran 100 MCMC iterations of the Bayesian competitors for settings where $p = 20^2$. For LASSO, we used the \verb|glmnet| \verb|R| package to implement ten-fold cross-validation (CV) to select parameters requiring tuning. For HBART, we used the \verb|rbart| \verb|R| package with models on the mean and variance as well as default parameters. We specified the Bayesian model with a horseshoe prior using the \verb|horseshoe| \verb|R| package. For the EBREG approach, we used the \verb|ebreg| \verb|R| package. We considered comparisons with Daye et al\cite{Daye2012}, but the computation requirements were prohibitive, with many settings running for over one hour per iteration.

\begin{figure}[t]
\centering
\includegraphics[width=5.5in]{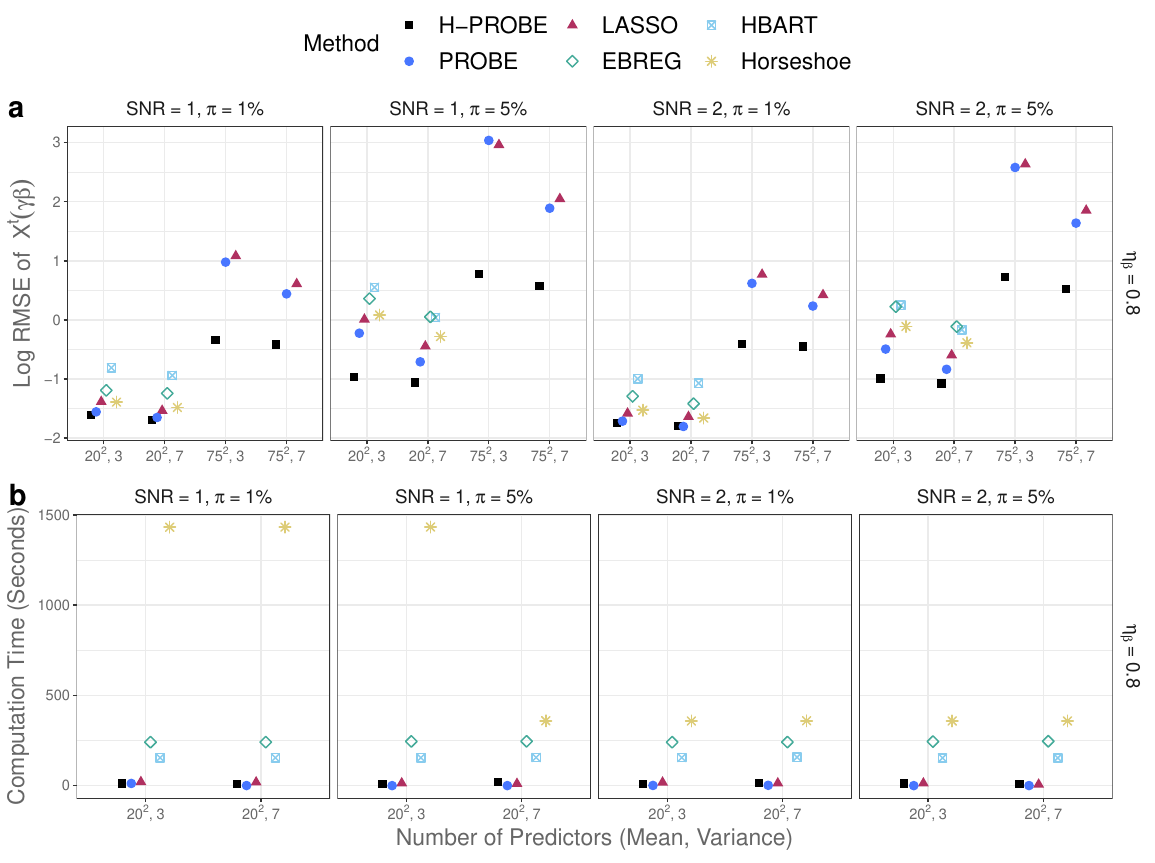} \\
\caption{(a) Log Root Mean Squared Errors (RMSE) of $\X' (\bgamma \be)$ and (b) Computation Time for H-PROBE (black squares), PROBE (blue circles), LASSO (maroon triangles),  EBREG (green diamonds), HBART (blue squares with inner cross), or a Bayesian model with a horseshoe prior (yellow stars), for selected simulation settings. \label{fig.rmse}}
\end{figure}

The LASSO, PROBE, EBREG, and horseshoe approaches do not model heteroscedasticity, while H-PROBE and HBART do. We compared the performance of the methods with Root Mean Squared Error (RMSE) and Median Absolute Deviation (MAD, in the Additional Simulation Results Section of Supplementary Materials) of $\X' (\bgamma \be)$, where data $\X$ consists of new observations not used during estimation (test set).  Figures \ref{fig.rmse} and in Supplementary Materials show that for nearly all simulation settings, H-PROBE had the lowest RMSE and MAD. The horsehoe approach had lower MAD than H-PROBE when $\pi = 1\%$. H-PROBE led to marked efficiency gains when the proportion of signals, the number of predictors on the mean, or the effect size of $\be$ were higher. EBREG, horsehoe, and HBART methods had computation time that was at the minimum eight-fold that of H-PROBE, PROBE, or LASSO in the smallest $p$ settings ($p = 20^2$). In the settings with the largest time difference, computation time grew to 18-fold the computational time of H-PROBE.

\begin{figure}[t]
\centering
\includegraphics[width=5.5in]{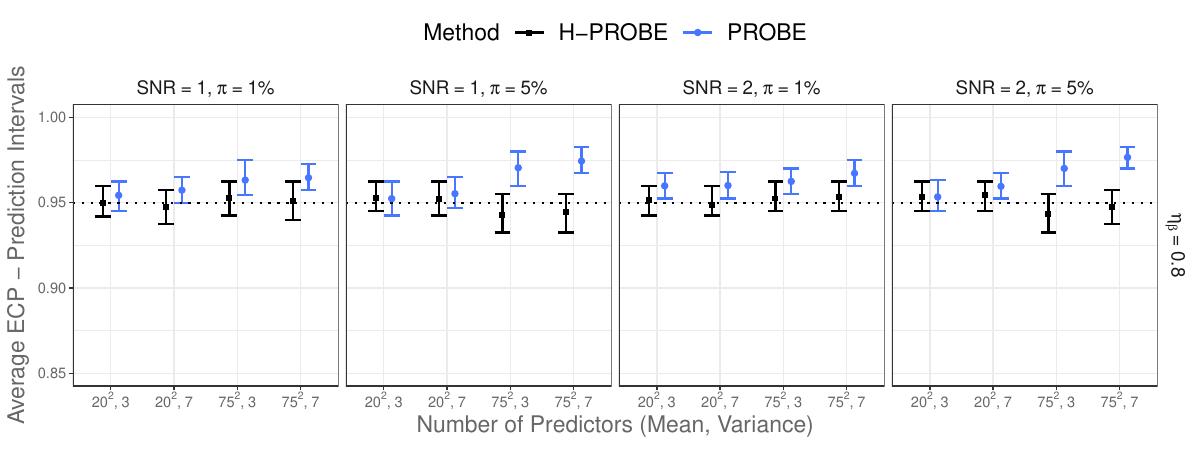} \\
\caption{Empirical Coverage Probabilities (ECPs) of Prediction Intervals (PIs) for $Y_{i,test}$ for H-PROBE (black squares) and PROBE (blue circles), for selected simulation settings. Vertical lines represent the first and third quartiles of the distributions of ECPs for PIs. \label{fig.ecppi}}
\end{figure}

For H-PROBE and PROBE, we also obtained empirical coverage probabilities (ECPs) of 95\% prediction intervals (PIs), which we estimated as the proportion of PIs that contained $Y_{i,test}$. Figure \ref{fig.ecppi} shows that the average ECP for H-PROBE PIs consistently remains centered at 0.95, whereas the average ECP for PROBE PIs exceeds 0.95 and has interquartile ranges that do not cover the 0.95 level, particularly when $p=75^2$. Supplementary Materials include a figure showing (log) PI lengths associated with the PI ECPs in Figure \ref{fig.ecppi}.
 
\begin{figure}[t]
\centering
\includegraphics[width=5.5in]{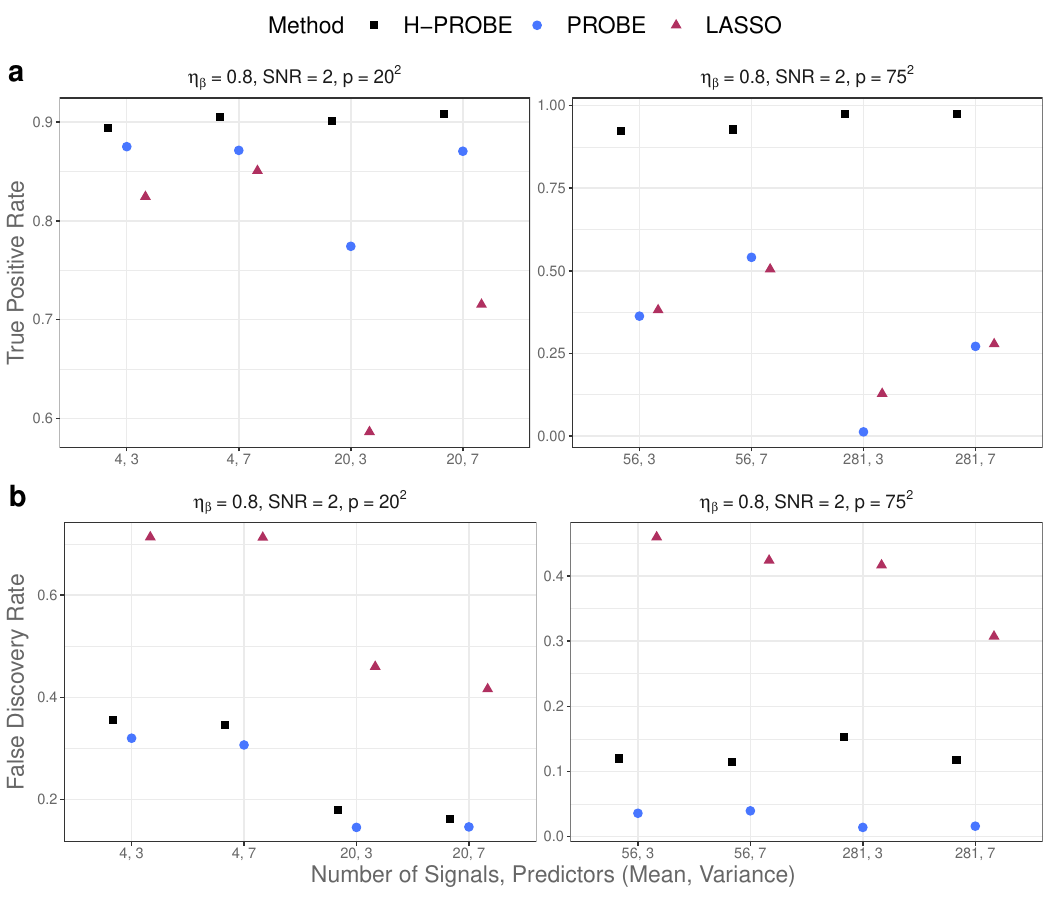} \\
\caption{True Positive Rate and False Discovery Rate for H-PROBE (black squares), PROBE (blue circles), or LASSO (maroon triangles), for selected simulation settings. \label{fig.select}}
\end{figure}

To compare the variable selection abilities of the methods, we calculated True Positive Rate (TPR) $TPR = \sum_{k;\gamma_k=1} \hat \gamma_k /|\bgamma|$ and the False Discovery Rate (FDR) $FDR = \sum_{k;\gamma_k=0} \hat \gamma_k /|\hat \bgamma|$ where $\hat \gamma_k=1$ if variable $k$ was `selected' for the given method. For H-PROBE and PROBE $\hat \gamma_k = I(p_k>0.5)$, while a predictor was selected for LASSO if the estimated coefficient was non-zero. H-PROBE performed well in variable selection. Figure \ref{fig.select}a shows that H-PROBE correctly selects the highest proportion of the true signals in all settings. This is strongly emphasized when the number of true signals ($|\bgamma|$) was $281$ with $p = 75^2$, and performance is closer between methods when there are only $4$ signals. In Figure \ref{fig.select}b, H-PROBE has a lower FDR than LASSO in all settings. Comparing the FDR between PROBE and H-PROBE, we see they are similar for $p=20^2$ and lower for PROBE when $p=75^2$ (where PROBE is markedly conservative).

\section{Data Analysis}\label{sec.da}

We return to our motivating example to illustrate the use and distinctive features of the H-PROBE method. While several researchers have predicted aphasia severity or type using brain images \citep{Yourganov2015, Lee2021, Teghipco2023}, none of these works addressed the fundamental issue of \emph{uncertainty quantification}. Point estimate predictions are often insufficient in clinical settings. Uncertainty quantification is crucial in precision medicine because it allows providers to assess the reliability of predictions and formulate optimal treatment plans.\citep{Begoli2019, Zou2023} Our application aims to use patients' imaging and brain damage data to predict \emph{and} quantify uncertainty of AQ score, which in turn guides post-stroke aphasia treatment decisions. 

The data include $n=167$ patients who have recently experienced a left-hemispheric stroke and are candidates for language rehabilitation therapy.\citep{Johnson2019, Yourganov2015} All individuals were scanned using a 3T MRI scanner, and an expert identified the lesion boundaries by hand via a high-resolution T2 scan. The lesions were then coregistered to the individual's T1 scan and warped to have a common size and shape through an enantiomorphic normalization clinical toolbox.\citep{Nacetal08, Roretal12} The resulting data contain over $5 \times 10^5$ binary features capturing the presence or absence of lesions in each 1 $mm^3$ brain voxel. Before comparing the performance of heteroscedastic and homoscedastic approaches in this clinical scenario, we streamlined the analysis by performing a marginal screening procedure based on Wang et al\cite{Wang2016} and retained $3 \times 10^4$ candidate imaging predictors. Total Brain Damage (TBD) was retained, i.e., the total number of voxels with lesions in the brain (out of $5 \times 10^5$). 

The relationship between AQ and TBD is displayed in Figure \ref{fig.intro1}a. Most AQ values are concentrated near the top of the range for lower brain damage and diffuse as brain damage increases. The residuals (squared and log-transformed) from homoscedastic PROBE by TBD are shown in Figure \ref{fig.intro1}b. There is a strong non-linear relationship between the error variance and TBD. Transformations of AQ also show evidence heterogeneity (see Figure \ref{fig.intro2} and transformation details in Supplementary Materials). As a result, we analyze AQ on the original scale and account for heteroscedasticity in this application. We use the H-PROBE approach, where the model on the variance includes TBD as well as its square-root transformation as predictors
\begin{eqnarray*}
\bSigma = diag(\exp \left\{- \left (\omega_1 \mbf{1} + \omega_2 \times \textbf{TBD} \ + \omega_3 \times \sqrt{\textbf{TBD}} \right ) \right\}. 
\end{eqnarray*}

Our analyses include Conformal Inference based on LASSO (\textit{split} variant) \citep{TibFoy19}, PROBE, and EBREG, a Bayesian method based on empirical priors for prediction in sparse high-dimensional linear regression.\cite{MarTan20} EBREG provides a simple algorithm based on a local search improvement rule that correctly identifies the support of the regression coefficients and subsequently provides PIs. We used \verb|R| packages \verb|probe|, \verb|conformalInference|, and \verb|ebreg| with 5-fold CV and default parameters.\citep{ebreg, TibFoy19, McLZgo21} For all methods, we modeled the mean of AQ ($\Y$) using the $3 \times 10^4$ candidate imaging predictors ($\X$), $\Y = \be \X + \mbf{e}$. Due to the high computational demands of fitting horseshoe and HBART when $p = 3 \times 10^4$, we omitted them from our analysis. 

To evaluate and compare the methods, we used 5-fold CV to calculate Mean Squared Predictive Error (MSPE), Median Absolute Deviation (MAD), and empirical coverage probability (ECP) of 95\% PIs where coverage implies that the PI for a given \textit{test} subject included the actual observation. Figures in the Additional Data Analysis Results Section of the Supplementary Materials provide additional results.

\begin{table}[t]
\caption{\label{table.brain} Performance metrics for H-PROBE and three comparison methods, PROBE, Conformal Inference, and EBREG, for the brain study analysis.}
\begin{center}
\begin{tabular}{c|c c c c} 
 \toprule
\textbf{Method} \textbackslash \textbf{Metric} & MAD & MSPE & Average PI Length & ECP\\  
 \midrule\\
\addlinespace[-2ex]
 H-PROBE & 8.635 & 220.671 & 60.749 & 0.952  
 \\ \addlinespace[0.8ex]
 \hline 
 \addlinespace[0.8ex]
 PROBE & 11.929 & 380.673 & 71.558 & 0.922 
 \\ \addlinespace[0.8ex]
 \hline 
 \addlinespace[0.8ex]
  Conformal Inference & 13.393 & 509.903 & 100.192 & 0.946 
 \\ \addlinespace[0.8ex]
 \hline 
 \addlinespace[0.8ex]
 EBREG & 17.677 & 614.314 & 91.582 & 0.940 
 \\ \addlinespace[0.8ex]
 \bottomrule 
\end{tabular} 
\end{center}
\end{table}

\begin{figure}[t]
\centering
\includegraphics[width=5.5in]{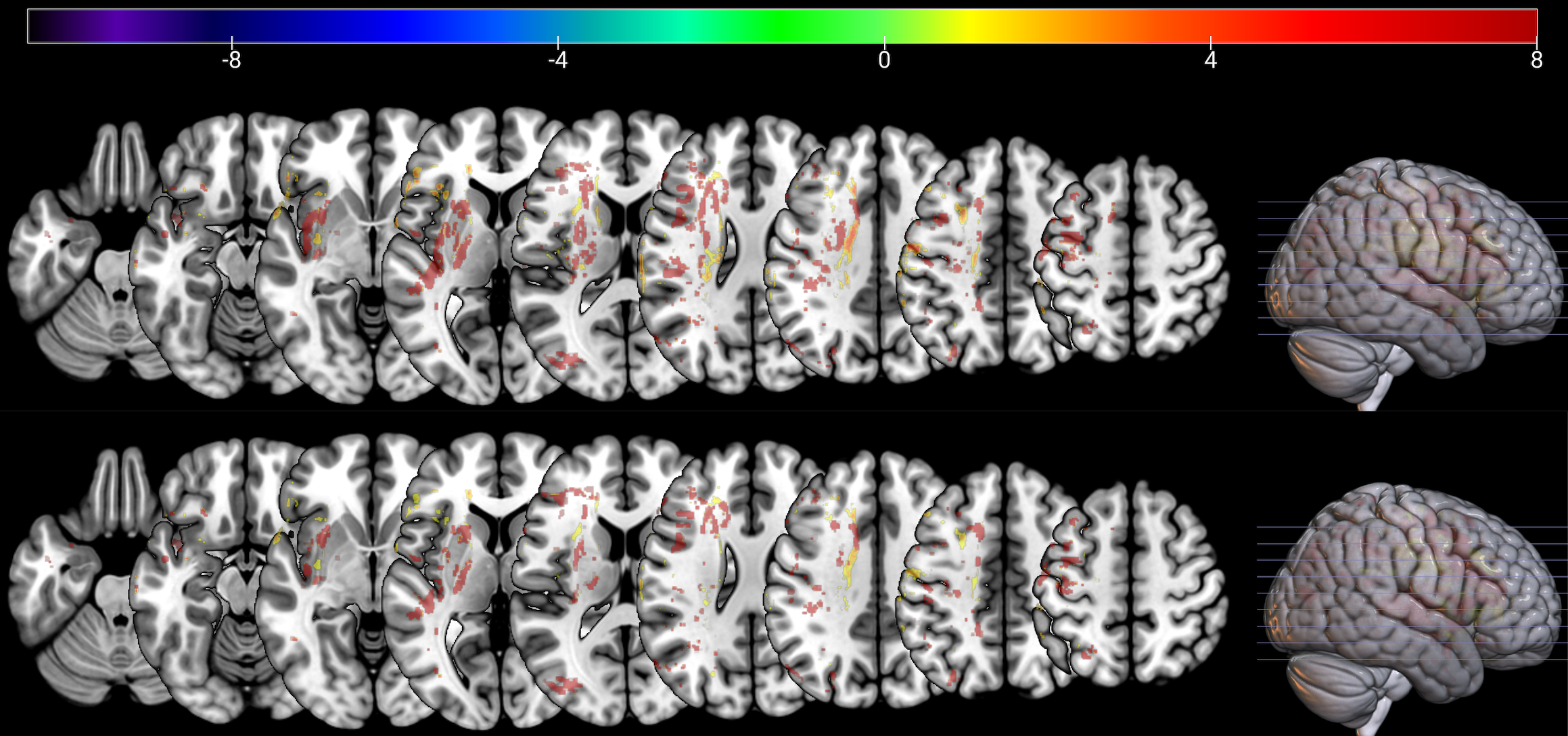} \\
\caption{Brain maps showing position, direction, and magnitude of voxel-specific $\be$ coefficients across different slides for PROBE (top) and H-PROBE (bottom). The color legend represents the magnitude of $\be$ estimates. \label{fig.brainmap}}
\end{figure}

Table \ref{table.brain} shows that H-PROBE had the lowest MAD and MSPE, followed by PROBE, Conformal Inference, and EBREG. H-PROBE also had the shortest average PI length with ECP close to the nominal $0.95$ level. This pattern is consistent with H-PROBE providing accurate predictions for new observations and down-weighting observations with high estimated variance. Figure \ref{fig.brainmap} provides per-voxel statistical brain maps that compare the performance of H-PROBE to PROBE. There is a large overlap between the voxels with positive $\be$ estimates between the two approaches. However, H-PROBE provided a more sparse model. This may be due to PROBE's misspecified homoscedastic model leading to more false discoveries. 

For both approaches, the voxels with large coefficients appear to be located in and around the Inferior Frontal Gyrus, which contains Broca's region, an important area of the brain linked to speech production. The pathophysiology of strokes results in detrimental effects on cerebral structures and functions. As a result, negative $\be$ estimates, which suggest a protective or beneficial impact of strokes on the brain, contradicts established neuroscientific understanding. As can be seen in Figure \ref{fig.brainmap}, most of the selected $\be$ estimates for H-PROBE and PROBE are positive. The LASSO model, which was markedly more sparse, resulted in only negative $\be$ estimates (brain maps are shown in Supplementary Materials). Among voxels with larger $\tilde{p}_k$ ($\tilde{p}_k>0.05$) all coefficients estimated by H-PROBE were positive, versus all negative for LASSO.

\begin{figure}[t]
\centering
\includegraphics[width=5.5in]{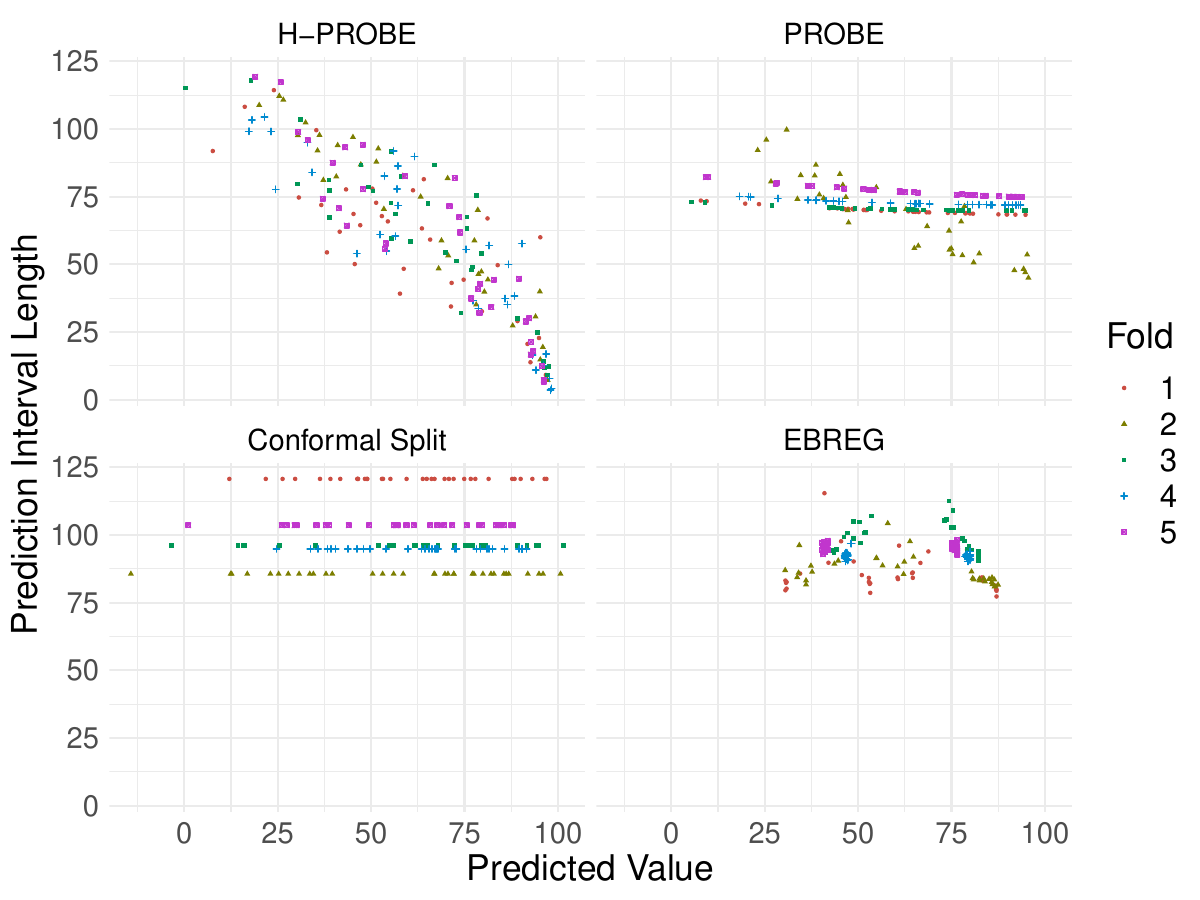} 
\caption{Prediction interval lengths by predicted values using H-PROBE and three comparison methods, PROBE, Conformal Inference, and EBREG. The color legend represents the cross-validation fold in which predictions were obtained. \label{fig.pi.preds}}
\end{figure}

\begin{figure}[t]
\centering
\includegraphics[width=5.5in]{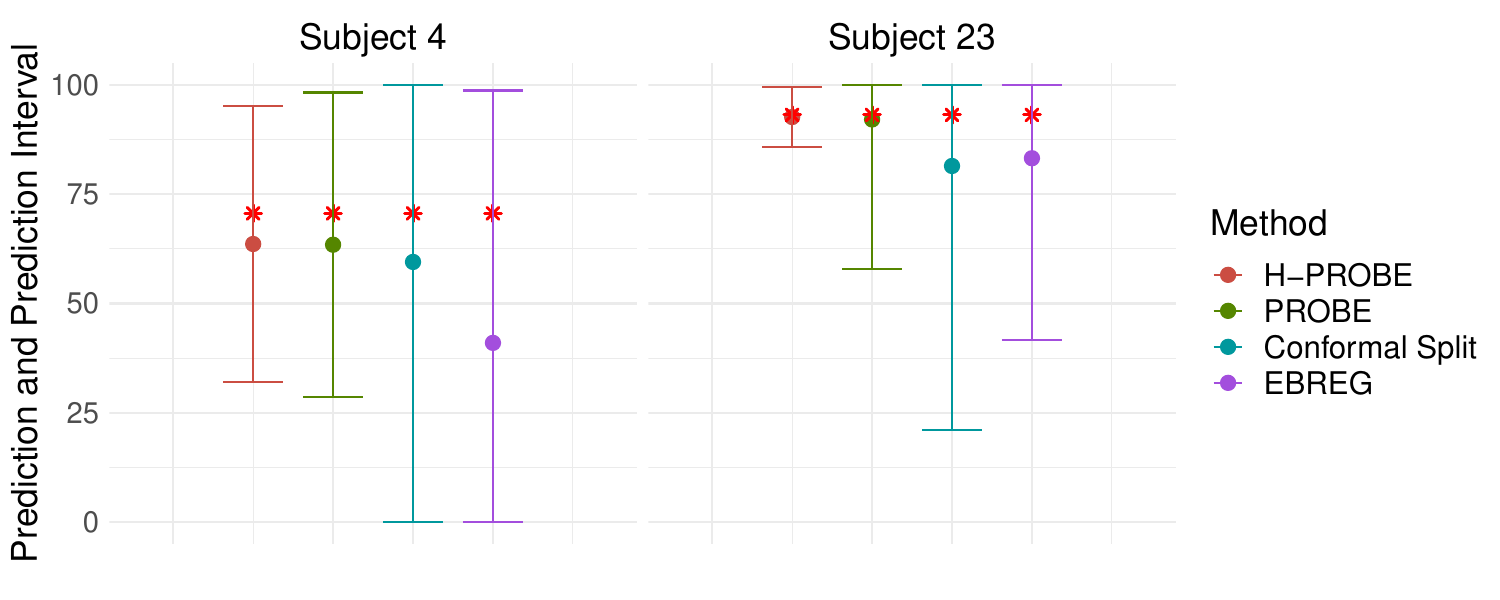} 
\caption{Predictions with prediction intervals using H-PROBE and three comparison methods, PROBE, Conformal Inference, and EBREG, for two random selected subjects. Red stars indicate the subject's true Aphasia Quotient score, while circles indicate predictions. Aphasia Quotient ranges that indicate aphasia severity are: AQ less than 26 is very severe, 26--50 is severe, 51--75 is moderate, and above 75 is mild. \label{fig.preds.subjects}}
\end{figure}

Figure \ref{fig.pi.preds} shows PI lengths for each patient and each method by fold. The Conformal Split and EBREG methods had similar PI lengths. As anticipated, H-PROBE displayed the widest range of PI lengths, reflected by the differing estimated $\tilde{\sigma}^2_i$ by subject (see $\tilde{\sigma}^2_i$ Figure in the Supplementary Material). While H-PROBE had the lowest average PI length across observations, it had the largest ECP, so the narrower PIs did not sacrifice coverage. We show predictions and PIs for two subjects from the study data in Figure \ref{fig.preds.subjects}. The panel for Subject 4 shows moderately wide PIs for H-PROBE and wider PIs for other methods, spanning multiple AQ severity levels. The AQ predictions for Subject 23 are mild using all methods, but only H-PROBE provided a tight PI that does not go below a the mild aphasia severity threshold. 

Shorter PIs impact the judgment of clinical outcomes like aphasia severity category (AQ less than 26 is very severe, 26--50 is severe, 51--75 is moderate, and above 75 is mild).\citep{Kertesz2007} For example, H-PROBE was the only method with PIs that spanned one aphasia severity category, where 29 subjects were predicted to have mild aphasia with the lower limit of their PIs above 75. None of the other methods provided PIs that spanned only one category. Conformal Split, EBREG, PROBE, and H-PROBE had $100\%$, $96\%$, $77\%$, and $63\%$ of their PIs cover at least three severity categories (out of four), respectively. Large uncertainty hinders the utility of Conformal Split, EBREG, and PROBE in clinical scenarios.

\section{Discussion}\label{sec.discuss}

In this paper, we have made both methodological and applied contributions. On the methodological side, we developed a novel approach to conduct high-dimensional linear regression for heteroscedastic data. H-PROBE uses a Bayesian framework with parameter expansion and minimally informative priors on the parameters. H-PROBE is a computationally effective solution to sparse linear regression in heteroscedastic settings that combines an empirical Bayes estimator with the PX-ECM algorithm. Simulation studies illustrated that accounting for heterogeneity in variance errors via the H-PROBE method generally resulted in more accurate estimation and prediction, as shown by lower MSEs and MADs for model predictions, compared to the PROBE, LASSO, HBART, EBREG, and horseshoe approaches. Empirical coverage probabilities of prediction intervals reached and stayed at the nominal 95\% level, with tighter PI lengths than other methods. H-PROBE is one of the only Bayesian approaches to address these issues in high-dimensional settings by including models on both the mean and the variance. Compared to HBART, H-PROBE is more scalable and leads to a parsimonious model, enhancing H-PROBE's interpretability in clinical settings.

On the application side, we added to the literature on prediction of aphasia severity by constructing PIs for AQ scores in stroke patients. Our analyses reinforced that appropriately accounting for non-constant error variances can lead to improvements in predictive ability and PI lengths while maintaining coverage. This has important clinical implications. When using markers such as predicted AQ to formulate treatment plans for post-stroke aphasia, not only is prediction accuracy important but so is providing measures of prediction uncertainty such as PIs and ECP.\citep{Begoli2019, Zou2023} 
H-PROBE provided narrow PIs where other methods could not (in patients with mild aphasia, AQ $> 75$), which allows clinicians to define treatment courses that weigh a relevant range of potential treatment outcomes a patient may experience based on the PIs of the predicted AQ. PIs and PI lengths that capture patient-level heteroscedasticity enable personalized treatment decisions based on each patient's stroke-induced total brain damage and MRI imaging. In this application, homoscedastic methods such as Conformal Inference, PROBE, and EBREG are likely to lead to less accurate predicted AQ and, thus, suboptimal treatment decisions compared to H-PROBE.

We focused our research on the situation where known markers of heterogeneity exist in the data.  However, in practice, the variables' impact on heterogeneity may be unknown. A valuable extension of H-PROBE is to the situation where model selection on mean and variance parameters is required.\citep{Chiou2020,Zhoetal21} If the heterogeneity markers are appropriately included in the variance model, observations with high residual variability will be accordingly down-weighted. 
The development of Bayesian high-dimensional variable selection methods that trim -- and possibly incorporate variables related to trimming (e.g., study sample size) -- as well as parametric models on the variance with transformations other than log are other worthy avenues of future research. 

A limitation of H-PROBE is that it assumes a linear model which may be insufficiently flexible. In contrast, HBART models both the mean and the variance nonparametrically.\citep{HBART2020} As a trade-off, however, HBART is much more computationally demanding than H-PROBE. One way to achieve greater flexibility for H-PROBE while retaining its computational advantages over HBART is to use a semiparametric additive model $Y_i = \sum_{k=1}^{p} f(X_{ik}) + \epsilon_i$. In this case, we could approximate each $f(X_{ik})$ as a linear combination of basis functions and use H-PROBE to regularize basis coefficients to zero, similar to Bai et al \citep{Bai2022} and Guo et al.\citep{GuoSIM2022} This is a useful extension for future work.

\bmsection*{Supplementary Materials}
The Supplementary Materials contain an expanded Methods section as well as additional results from simulations and real data analyses. (.pdf)

\bibliography{Dissertation_References}

\pagebreak

\setcounter{section}{0}
\setcounter{figure}{0}   
\setcounter{table}{0}    
\renewcommand{\thesection}{\Alph{section}}
\renewcommand\thefigure{\thesection.\arabic{figure}}    
\renewcommand\thetable{\thesection.\arabic{table}}    

\begin{center}
\huge{SUPPLEMENTARY MATERIALS \\ for \\ Quantifying predictive uncertainty of aphasia severity in stroke patients with sparse heteroscedastic Bayesian high-dimensional regression}
\end{center}

\bigskip

\bmsection{Expanded Methods Section}\label{sec.exp.methods}
\vspace*{12pt}

\bmsubsection{Model framework}\label{sec.blmm.setup.sm}

The complete model framework is presented in the main text. We briefly review the models on the mean and the variance. 
Our model on the mean allows for sparse and non-sparse predictors. For ease of presentation, we assume the mean's non-sparse predictors are also the variance predictors, denoted by $\V \in \mathcal{R}^{n \times v}$. In practice, this need not be the case.  We write the model on the mean as 
\begin{equation}\label{eq.mod.sm}
\Y = \X (\bgamma \be) + \V \bvarphi + \mbf{\epsilon},
\end{equation}
where $\bgamma \be$ is a Hadamard product, $\bgamma \in \{0,1\}^p$, and $\bvarphi \in \mathcal{R}^{v}$. Let $\mbf{\mathcal{D}} = \{\mathcal{D}_1,\ldots,\mathcal{D}_n\}$ with $\mathcal{D}_i = (Y_i,\X_i,\V_i)$ denote the observed data. 

We add the following parametric model on the diagonal variance matrix $\bSigma$,
\begin{eqnarray}\label{eq.modvar.sm}
-\log\{diag(\bSigma)\} &=& \V\bomega, 
\end{eqnarray}
where  $\bomega \in \mathcal{R}^{v}$. The log transformation on the variances ensures positivity, can accommodate variances that vary over orders of magnitude, and has been long established in variance function modeling.\citep{Carroll1988, Cleveland1993}

We use an EM algorithm for estimation. Within the algorithm, the parameter estimates at iteration $t$ are used to obtain the expected complete-data log-posterior distribution with respect to $\bgamma$,
\begin{equation}\nonumber
E_{\bgamma}\left\{ \log \ p( \be, \bvarphi, \bomega | \mbf{\mathcal{D}}, \bgamma)| \mbf{\mathcal{D}}, \be^{(t)}, \bvarphi^{(t)}, \bomega^{(t)} \right\},
\end{equation}
which is sequentially maximized to obtain MAP estimates $\be^{(t+1)}$, $\bvarphi^{(t+1)}$, and $\bomega^{(t+1)}$ (i.e., $\bSigma^{(t+1)}$) (discussed in Sections \ref{sec.first.m.sm} and \ref{sec.first.e.sm}). Finally, for $\gamma_k$, plug-in empirical Bayes estimators of $\pi$ and $f_{\beta}$ are used to estimate the posterior expectation $\mbf{p} = (p_1, \ldots, p_p)$ where $p_k = P(\gamma_k = 1 | \mbf{\mathcal{D}}, \pi)$ (discussed in Section \ref{sec.first.e.sm}).

\bmsubsection{Overview and notation}\label{sec.part.setup.sm}

We begin this Section with an overview of the PROBE algorithm.\citep{McLain2022} The aim is to perform MAP estimation for $\bgamma$, $\be$, and $\bvarphi$ in a high-dimensional setting. PROBE is based on a quasi Parameter-Expanded Expectation-Conditional-Maximization (PX-ECM) algorithm, which is a combination of the ECM  and PX-EM algorithms.\citep{Meng1993,Liuetal98} The CM-step of the PX-ECM results in coordinate-wise (predictor-specific) optimization, where the remaining parameters are restricted to their values at the previous iteration. As a result, both PROBE and H-PROBE partition the mean of the model (\ref{eq.mod.sm}) by predictor $k$. The notation $\bA_{\char`\\ k}$ indicates the matrix, vector, or collection without predictor or element $k$. Define $\W_k = \X_{\char`\\ k}(\bgamma_{\char`\\ k} \be_{\char`\\ k}) + \V\bvarphi = (W_{1k}, \ldots, W_{nk})'$ and thus  
%
$E(\Y|\W_k) =   \X_k \beta_k + \W_k,$
%
where $\W_k$ encompasses the impact of all predictors except $\X_k$. 

Since $\W_k$ is estimated as part of the algorithm, we use parameter-expansion \citep{Liuetal98} to include variables $\alpha_k$, which adjust for the impact of $\W_k$ when updating $\beta_k$ for all $k=1,\ldots,p$. This results in
\begin{eqnarray}\label{eq.expand.part.sm}
E(\Y|\W_k) =   \X_k \beta_k + \W_k \alpha_k,
\end{eqnarray}
for $k=1,\ldots,p$. The $\alpha_k$ parameter helps estimate the posterior variance of $\beta_k|(\gamma_k=1)$ more accurately since it accounts for the dependence between $\W_k$ and $\X_k$ (discussed further in Section \ref{sec.first.m.sm}). This posterior variance is required in our implementation of the E-step and used to create prediction intervals. 
In H-PROBE, the goal is to estimate the MAP of $\bvarphi$, $\bomega$, and then $\beta_k|(\gamma_k=1)$ for all $k=1,\ldots,p$ in sequence.  Table \ref{methods.table.est.sm} provides a summary of the parameters estimated by H-PROBE.

\renewcommand{\arraystretch}{1.3}
\begin{table}[t]
\caption{Parameters and estimates provided by H-PROBE. SM = Supplementary Materials.}
\label{methods.table.est.sm}
\begin{center}
\begin{tabular}{||p{1.3cm} p{1.2cm} p{1.2cm} p{8cm} ||} 
 \hline
 Parameter & Estimate & Equation & Parameter Definition \\ 
 \hline\hline 
 \space $\be$ & $\tilde{\be}$ & (\ref{eq.mod.sm}), (\ref{eq.expand.part.sm}) & Vector of sparse regression coefficients for predictors in $\X$, conditional on $\bgamma = 1$, in the model on the mean. \\
 \hline
 $\mbf{S}^2$ & $\tilde{\mbf{S}}^2$ & SM & Vector of posterior variances of $\be|(\bgamma = 1)$. \\ 
 \hline
 $\bgamma$ & $\tilde{\mbf{p}}$ & (\ref{eq.mod.sm}), (\ref{eq.expand.part.sm}) & Vector of inclusion indicators for sparse coefficients $\be$ associated with predictors in $\X$. \\
 \hline
 $\bvarphi$ & $\tilde{\bvarphi}$ & (\ref{eq.mod.sm}) & Vector of non-sparse regression coefficients in the model on the mean. \\
 \hline
 $\W_0$ & $\tilde{\W}_0$ & (\ref{eq.expand.part.sm}) & Overall (non-partitioned) latent parameter $\W_0 = \X(\bgamma \be)$. \\
 \hline
 $\alpha$ & $\tilde{\alpha}$ & (\ref{eq.expand.part.sm}) & Overall (non-partitioned) coefficient adjusting for the impact of overall (non-partitioned) $\W_0$. \\
 \hline
 $\bomega$ & $\tilde{\bomega}$ & (\ref{eq.modvar.sm}) & Vector of non-sparse regression coefficients in the model on the variance. \\ 
 \hline
 $\bSigma$ & $\tilde{\bSigma}$ & (\ref{eq.modvar.sm}) & Diagonal variance matrix $\bSigma = \exp\{-(\V \bomega)\}$. \\ 
 \hline
\end{tabular}
\end{center}
\end{table}

Note that $\W_k$ is unknown since it is a function of parameters $(\be_{\char`\\ k},\bvarphi)$ and the ``missing data'' $\bgamma_{\char`\\ k}$. In the CM-steps discussed in Section \ref{sec.first.m.sm}, $\bvarphi$ and $\be_{\char`\\ k}$ are fixed to their values from the previous iteration when estimating $\beta_k$, for all $k$.\citep{Meng1993} 
The E-step discussed in Section \ref{sec.first.e.sm} updates $E(\W_k)$ and $E(\W^2_k)$, where the expectations are over $\bgamma_{\char`\\ k}$. After the moments of the $\W_k$'s are updated, they are used to perform the subsequent CM-steps. The MAP of $\bomega$ does not have a closed form. As a result, we perform the maximization for this parameter via quasi-Newton optimization.\citep{Fletcher1987} More discussion contrasting PROBE to other extensions of the EM is available in McLain et al.\cite{McLain2022}

Along with $\W_{k}$ and $\Z_{k} = (\X_k \ \W_k)$ described above, we define an `overall' non-partitioned $\W$,  denoted by $\W_0 = \X (\bgamma \be)$, and $\Z_{0} = (\V \ \W_0)$. The calculations in Section \ref{sec.first.e.sm} require $W_{i0}^{(t-1)} = E(W_{i0}|\be^{(t-1)}, \mbf{p}^{(t-1)})$, \\ $W_{ik}^{(t-1)} = E(W_{ik}|\be_{\char`\\ k}^{(t-1)}, \mbf{p}_{\char`\\ k}^{(t-1)}, \alpha_0^{(t-1)}, \bomega^{(t-1)}, \bvarphi^{(t-1)})$ for $k\geq 1$, and \\ $\W_{\ell}^{(t-1)} = (W_{1\ell}^{(t-1)}, \ldots, W_{n \ell}^{(t-1)})$ for $\ell \in (0,\ldots,p)$, with analogous notation for the second moments $W_{i0}^{2(t-1)}$, $W_{ik}^{2(t-1)}$, and $\W_{ \ell}^{2(t-1)}$. Some key quantities in the CM-step updates are $(\W^{'}_{\ell} \bSigma^{-1})^{(t-1)} = \sum_{i=1}^{n} W_{i \ell}^{(t-1)}/\sigma_i^{2(t-1)}$ and $(\W^{'}_{\ell} \bSigma^{-1} \W_{\ell})^{(t-1)} =\sum_{i=1}^{n} W_{i \ell}^{2(t-1)}/\sigma_i^{2(t-1)}$ where $\sigma_i^{2(t-1)} = \exp(-\V_i \bomega^{(t-1)})$. Further, let
\begin{eqnarray}\label{xtx}
(\Z_{k}^{\prime} \bSigma^{-1} \Z_{k})^{(t-1)} = \left(
\begin{array}{cc}
\X'_{k} \bSigma^{-1(t-1)} \X_{k}  & \X_{k}^{\prime} (\bSigma^{-1} \W_{k})^{(t-1)} \\
(\W^{'}_{k} \bSigma^{-1})^{(t-1)} \X_{k} & (\W^{'}_{k} \bSigma^{-1} \W_{k})^{(t-1)}
\end{array}  \right) \quad \mbox{for }  \ k = 1,\ldots,p.
\end{eqnarray} 
We define $(\Z^{'}_{0} \bSigma^{-1} \Z_{0})^{(t-1)}$ similarly to (\ref{xtx}) where $\X_k$ and $\W_{k}$ are replaced with $\V$ and $\W_{0}$, respectively.

\bmsubsection{CM-step}\label{sec.first.m.sm}

The CM-steps maximize the expected complete-data log-posterior distribution of $\bfeta_{\ell}$ for each $\ell \in (0,1,\ldots,p)$. An `overall' model associated with the non-partitioned $\W_0$ is estimated when $\ell = 0$ via $E(\Y|\W_{0}) = \V\bvarphi + \alpha_0 \W_0$ and focuses on the parameters $\bfeta_0=(\bvarphi, \alpha_0, \bomega)$. The remaining partitions $\ell \ge 1$ focus on $\bfeta_k=(\beta_k, \alpha_k)'$ via (\ref{eq.expand.part.sm}). The complete-data log-posterior distribution is denoted by $l(\bfeta_{\ell}|\mbf{\mathcal{D}},\W_{\ell}, \mbf{\Gamma}_{\ell})$ and 
its expectation $E_{{\ell}}\{l(\bfeta_{\ell}|\Y,\W_{\ell}, \mbf{\Gamma}_{\ell}) \}$
is conditional on $\be_{\char`\\ k}$, $\mbf{p}_{\char`\\ k}$, and $\bfeta_0$ for all $\ell \geq 1$, and on $\be$ and $\mbf{p}$ for $\ell = 0$. 
At iteration $t$, the CM-step maximizes
\begin{equation}\label{eq.first.m.sm}
\hat{\bfeta}_{\ell}^{(t)} = \mbox{argmax}_{\bfeta_{\ell}}E^{(t-1)}_{{\ell}}\{l(\bfeta_{\ell}|\Y,\W_{\ell}, \mbf{\Gamma}_{\ell}) \} \ \mbox{for} \ \ell=0,1,\ldots,p,
\end{equation}
where $E^{(t-1)}_\ell$ represents the expectation given the relevant parameter estimates at iteration $(t-1)$, i.e., given ($\be,\mbf{p}$) for $\ell =0$ and ($\be_{\char`\\ \ell},\mbf{p}_{\char`\\ \ell},\bfeta_0$) for $\ell \geq 1$. $\mbf{\Gamma}_{\ell}$ represents the hyperparameters for $\bfeta_{\ell}$.

For the overall model (when $\ell = 0$) at iteration $t$, the MAP values for $(\bvarphi, \alpha_0)$ are
\begin{eqnarray}\label{eq.calib.sm}
\begin{pmatrix}
    \hat \varphi^{(t)} \\ \hat\alpha_0^{(t)}
\end{pmatrix}  = \left\{(\Z^{'}_{0} \bSigma^{-1} \Z_{0})^{(t-1)}\right\}^{-1} \left(\Z_{0}^{'} \bSigma^{-1}\right)^{(t-1)} \Y.
\end{eqnarray}
For $\bomega^{(t)}$, the MAP has no closed form. As a result, we use a quasi-Newton optimizer \citep{Fletcher1987} on the log posterior distribution of $\bomega$ to find the MAP estimate $\bomega^{(t)}$. Then, we obtain $ \bSigma^{-1(t)} = \exp\{-(\V' \bomega^{(t)})\}^{-1}$. For the $k$th partition of iteration $t$, the MAP of $\mbf{\xi}_k$ is
\begin{equation}\label{eq.beta.upd.sm}
\hat{\mbf{\xi}}^{(t)}_k=\left\{ (\Z_{k}' \bSigma^{-1} \Z_{k})^{(t-1)} \right\}^{-1} \left(\Z^{'}_{k} \bSigma^{-1}\right)^{(t-1)} \Y. 
\end{equation}

The E-step used in Section \ref{sec.first.e.sm} requires an estimate of the posterior variance of $\beta_k|(\gamma_k=1)$. Here, the posterior covariance of $\hat{\mbf{\xi}}^{(t)}_k$ is estimated by 
\begin{equation}\label{eq.cov.ksi.sm} 
\{(\Z_{k}'  \bSigma^{-1} \Z_{k})^{(t-1)}\}^{-1}\{ (\Z^{\prime}_{k}  \bSigma^{-1})^{(t-1)} \Z^{(t-1)}_{k}\}\{(\Z_{k}'  \bSigma^{-1} \Z_{k})^{(t-1)}\}^{-1},
\end{equation}
where $\hat S^{2(t)}_k$ denotes the $(1,1)$ element. 
We use the $\hat S^{2(t)}_k$'s 
to create prediction intervals.

\bmsubsection{E-step}\label{sec.first.e.sm}

Before commencing with the E-step, to accelerate convergence we limit the step size using learning rates $q^{(t)}$ via
\begin{eqnarray}\label{eq.mix.sm}
\begin{aligned}
\beta^{(t)}_k &= (1-q^{(t)})\beta^{(t-1)}_k  + q^{(t)}\hat{\beta}^{(t)}_k, \ \mbox{and} \\ 
S^{2(t)}_k &= \{(1-q^{(t)})(S^{2(t-1)}_k)^{-1} + q^{(t)}(\hat{S}_k^{2(t)})^{-1}\}^{-1}. 
\end{aligned}
\end{eqnarray}
We use $q^{(t)} = \frac{1}{t+1}$, which creates a moving average of $\beta_k$ over iterations. 
McLain et al\cite{McLain2022}, Minka et al\cite{MinLaf02}, and Vehtari et al\cite{Vehetal20}, and  provide a discussion on learning rates and their implications. 

To estimate $p_k = P\left(\gamma_k = 1| \Y, \pi_0 \right)$
while maintaining uninformative priors, we use a plug-in empirical Bayes estimator, which is motivated by two-groups approach to multiple testing where many test statistics with zero and non-zero expectations are available.\citep{CasRoq20, Efr08, Lietal08}
We define test statistics as $\mathcal{T}^{(t)}_k=\beta^{(t)}_k/S^{(t)}_k$ with distribution $(1-\gamma_k)f_{\mathcal{Z}}(\cdot) + \gamma_k f_1(\cdot)$, where $f_{\mathcal{Z}}(\cdot) \sim N(0,1)$ and $f_1$ is unknown and a function on $f_\beta$. We also require $\pi_0$, the proportion of null hypotheses. This yields the plug-in empirical Bayes estimator of the posterior expectation of $\gamma_k$ as
\begin{equation}\label{eq.p.sm}
p^{(t)}_k = 1-\frac{\hat \pi^{(t)}_0 f_0(\mathcal{T}_k^{(t)})}{\hat{f}^{(t)}_1(\mathcal{T}_k^{(t)})}, 
\end{equation}
where $\hat \pi^{(t)}_0$ and $\hat{f}^{(t)}_1$ are empirical Bayes estimates of $\pi_0$ and $f_1$ based on the observed $\mathcal{T}^{(t)}_k$'s. In our simulations and data analyses, we use $\hat{\pi}^{(t)} = \sum_k I({P}^{(t)}_k\geq \lambda)/\{p \times (1-\lambda)\}$, based on Storey \cite{Sto07}, where ${P}_k^{(t)}$ is a two-sided p-value for $\mathcal{T}^{(t)}_{k}$ and $\lambda=0.1$ \citep{Blanchard2009}, and Gaussian kernel density estimation on ${\mbf{\mathcal{T}}}^{(t)}= (\mathcal{T}^{(t)}_1,\ldots,\mathcal{T}^{(t)}_p)$ to obtain $\hat{f}^{(t)}$.\citep{Sil86} 

Finally, we estimate the first and second moments of $\W_{\ell}$. These moments are expectations of $\W_{\ell}$ over the unknown $\bgamma$. Through the use of the ECM, the values of $\be$ and $\varphi$ are fixed at their estimates from the previous iteration. Independence between $\gamma_k$'s allows effective computation to be performed at the observation $i$ level through
\begin{eqnarray}\label{eq.EW.sm}
W^{(t)}_{i 0} = E\{ \X_{i} (\bgamma\be) |\be^{(t)}, \mbf{p}^{(t)}\} = \X_{i} (\be^{(t)} \mbf{p}^{(t)} ),
\end{eqnarray}
and $W^{2(t)}_{i 0} = E(W_{i 0}^2|\be^{(t)}, \mbf{p}^{(t)}) = Var(W_{i0}|\be^{(t)}, \mbf{p}^{(t)}) + ( W^{(t)}_{i 0} )^2$ where 
\begin{eqnarray}\label{eq.VarW.sm}
Var(W_{i0}|\be^{(t)}, \mbf{p}^{(t)})  = \X_{i}^2 \left\{\be^{(t)2}  \mbf{p}^{(t)}(1- \mbf{p}^{(t)}) \right\}.
\end{eqnarray}
Online calculations of $\W_k^{(t)}$ and $\W_k^{2(t)}$ are made by subtracting the contributions of the $k$th predictor from $\W^{(t)}_0$ and $\W^{2(t)}_0$, respectively. As a result, the high-dimensional matrix computations in (\ref{eq.EW.sm}) and (\ref{eq.VarW.sm}) only need to be made once per iteration.

\bmsubsection{Model checks}\label{sec.conv.pred.sm}

The H-PROBE method converges when subsequent changes in $\W_0^{(t)}$ are small since $\W_0^{(t)}$ captures the trajectory of all model parameters. Specifically, convergence at iteration $t$ is quantified via \\ $CC^{(t)} = \log(n)\max_{i}\left\{(W^{(t)}_{i 0} - W_{i 0}^{(t-1)})^2/ Var(W_{i0}|\be^{(t)}, \mbf{p}^{(t)})\right\}$, where the ECM algorithm has converged when $CC^{(t)} < \chi^2_{1,0.1}$ and $\chi^2_{1,0.1}$ represents the $0.1$th quantile of a $\chi^2$ distribution with $1$ degree of freedom. 
We initiate the algorithm using $\be^{(0)} = \mbf{0}$ and $\mbf{p}^{(0)} = \mbf{0}$, which gives $\W^{(0)} = \mbf{0}$ and $\W^{2(0)} = \mbf{0}$. For the elements of $\bomega^{(0)}$, we initialize the first element to $\log(s^2_Y)$, where $s^2_Y$ is the sample variance of $\Y$, and all remaining elements to $0$. 

Algorithm \ref{algo.1.sm} shows H-PROBE steps in sequence. Upon convergence, H-PROBE provides MAP estimates $\tilde{\be}$, $\tilde{\mbf{p}}$, $\tilde{\bvarphi}$, $\tilde{\alpha}_0$, $\tilde{\bomega}$, $\tilde{\bSigma}$ as well as $\tilde{{S}}_k^2$, the posterior variance of $\tilde{\beta}_k|(\gamma_k=1)$, for all $k$.  
While the properties of $\tilde{\bvarphi}$ are not the focus of this research, we do wish to account for the uncertainty it contributes to the MAP estimates in prediction intervals. To this end, let $\tilde \Psi =  \{(\Z_{0}'  \tilde{\bSigma}^{-1} \Z_{0})^{(t-1)}\}^{-1}\{ (\Z^{\prime}_{0}  \tilde{\bSigma}^{-1})^{(t-1)} \Z^{(t-1)}_{0}\}\{(\Z_{0}'  \tilde{\bSigma}^{-1} \Z_{0})^{(t-1)}\}^{-1}$ denote the estimated posterior covariance of $(\tilde{\bvarphi},\tilde{\alpha}_0)$. Prediction intervals for a future observation are presented in detail in the main text. 

\begin{algorithm}[t]
\caption{The H-PROBE algorithm }\label{algo.1.sm}
\begin{algorithmic}[]
 \State  Initialize $\mbf{W}^{(0)}$, $\mbf{W}^{2(0)}$, and $\bSigma^{(0)}$
    \While{$CC^{(t)} \geq \chi^2_{1,\varepsilon} $ and $\max(\mbf{p}^{(t)})>0$} 
   \State  \textbf{CM-step}
        \State \hspace{0.6 cm} Use $\mbf{W}^{(t-1)}_\ell$ and $\mbf{W}^{2(t-1)}_\ell$ to estimate $\bfeta^{(t)}_{\ell}$ for $\ell=0,1,\ldots,p$ via (\ref{eq.calib.sm})-(\ref{eq.beta.upd.sm}).
\State \textbf{E-step}
        \State\hspace{0.6 cm} (a) Calculate $\beta^{(t)}_k$ and $S^{2(t)}_k$ using (\ref{eq.mix.sm}) for all $k$.
    \State\hspace{0.6 cm} (b) Estimate $\hat{f}^{(t)}$ and $\hat{\pi}^{(t)}_{0}$ and use them to calculate $\mbf{p}^{(t)}$ via (\ref{eq.p.sm}).
    \State\hspace{0.6 cm} (c) Calculate $\mbf{W}^{(t)}$ and $\mbf{W}^{2(t)}$ via (\ref{eq.EW.sm}) and (\ref{eq.VarW.sm}).
\State Calculate $CC^{(t)}$ and check convergence.
    \EndWhile  
\end{algorithmic}
\end{algorithm}

\begin{figure}[t]
\centering
\includegraphics[width=5.5in]{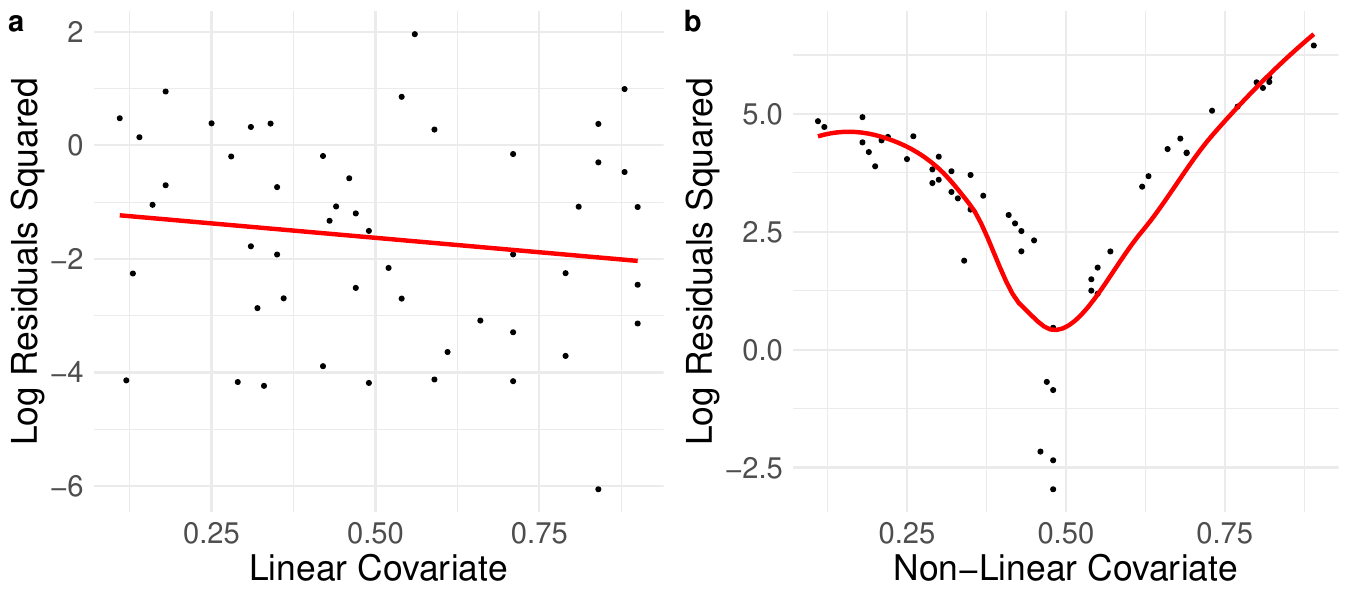} \\
\caption{Log Residuals Squared for homoscedastic LASSO models plotted against a linear and non-linear heterogeneity variable. \label{fig.assumptions.sm}}
\end{figure}

The residuals from a high-dimensional homoscedastic regression model can be used to identify variables that may have an association with the residual variance and appropriate forms of such variables such that the parametric assumptions of the variance model in (\ref{eq.modvar.sm}) are adequately met. Levene's, Barlett's, or Brown-Forsythe are tests that can be used to assess if the variance of the residuals is related to a candidate heterogeneity variable.\citep{SebLee03} Further, plots of the residuals or the log-squared residuals can be used to determine if transformations of variables are necessary. Figure \ref{fig.assumptions.sm} shows an example of residual plots used to validate model assumptions. Both panels are from homoscedastic LASSO models with $50$ observations of $100$ sparse predictors. The data used in each model was generated such that the residual variance is associated with a linear covariate (Panel (a)) and a non-linear covariate ($variable + variable^2$, Panel (b)). In both plots, the covariate has a relationship with the residual variance. This association is weaker in Panel (a), while it is stronger in Panel (b). The residuals in Panel (a) show that the linear form of the heterogeneity variable fulfills the parametric assumptions of the variance model in (\ref{eq.modvar.sm}), whereas Panel (b) shows that additional non-linear transformations of the heterogeneity variable are needed to fulfill the assumptions of the variance model. 

\bmsection{Additional Simulation Results}\label{sec.sm.sim}
\vspace*{12pt}

\begin{figure}[t]
\centering
\includegraphics[width=5.5in]{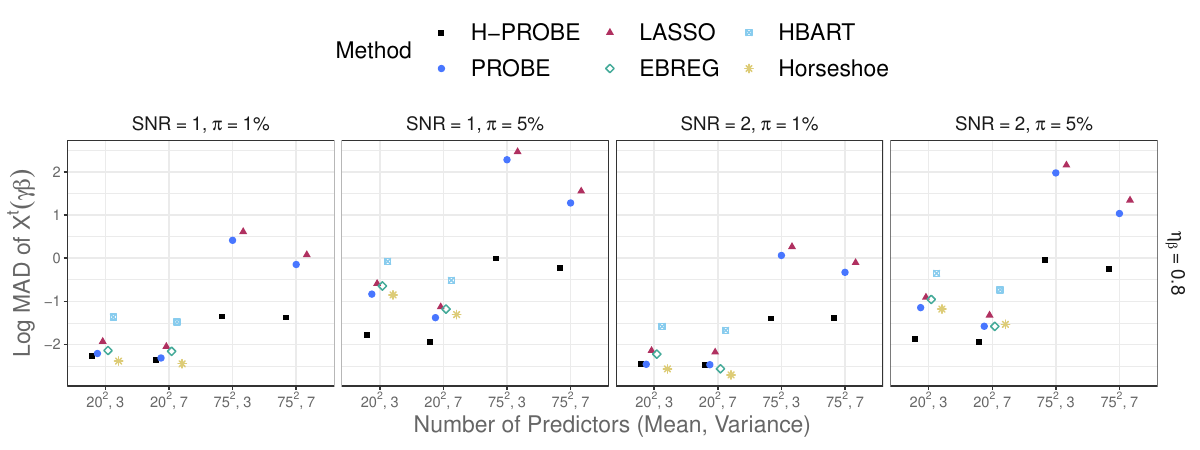} \\
\caption{Log Median Absolute Deviation (MAD) of $\X' (\bgamma \be)$ for H-PROBE (black squares), PROBE (blue circles), LASSO (maroon triangles),  EBREG (green diamonds), HBART (blue squares with inner cross), or a Bayesian model with a horseshoe prior (yellow stars), for selected simulation settings. \label{fig.mad}}
\end{figure}

We examined the performance of H-PROBE by focusing on the Root Mean Squared Error (RMSE, main text) and Median Absolute Deviation (MAD) of $\X' (\bgamma \be)$, where data $\X$ consists of new observations not used during estimation (test set). Figure \ref{fig.mad} shows that for all simulation settings except one, H-PROBE had the lowest MAD, especially when the proportion of signals, the number of predictors on the mean, or the effect size of $\be$ were higher. Results are only shown for $\eta_{\beta} = 0.8$ for brevity. In the setting where H-PROBE slightly underperformed compared to PROBE, the MAD was $6\%$ higher for H-PROBE.

\begin{figure}[t]
\centering
\includegraphics[width=5.5in]{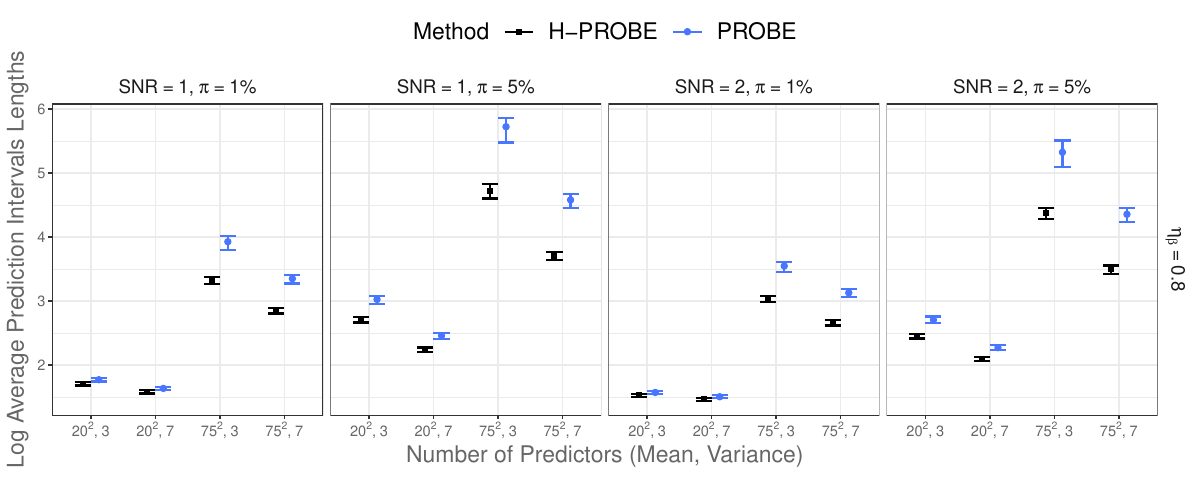} \\
\caption{Log Average Prediction Intervals (PI) lengths for $Y_{i,test}$ for H-PROBE (black squares) and PROBE (blue circles), for selected simulation settings. Vertical lines represent the first and third quartiles of the distributrions of PI lengths. \label{fig.pilen}}
\end{figure}

Figure \ref{fig.pilen} shows Prediction Interval (PI) lengths associated with the PI Empirical Coverage Probabilities (ECPs) in the main text.
The PI lengths in the settings where PI ECPs for the PROBE method exceeded 95\% were the widest, however, those ECPs did not actually cover the 0.95 level.

\begin{figure}[t]
\centering
\includegraphics[width=5.5in]{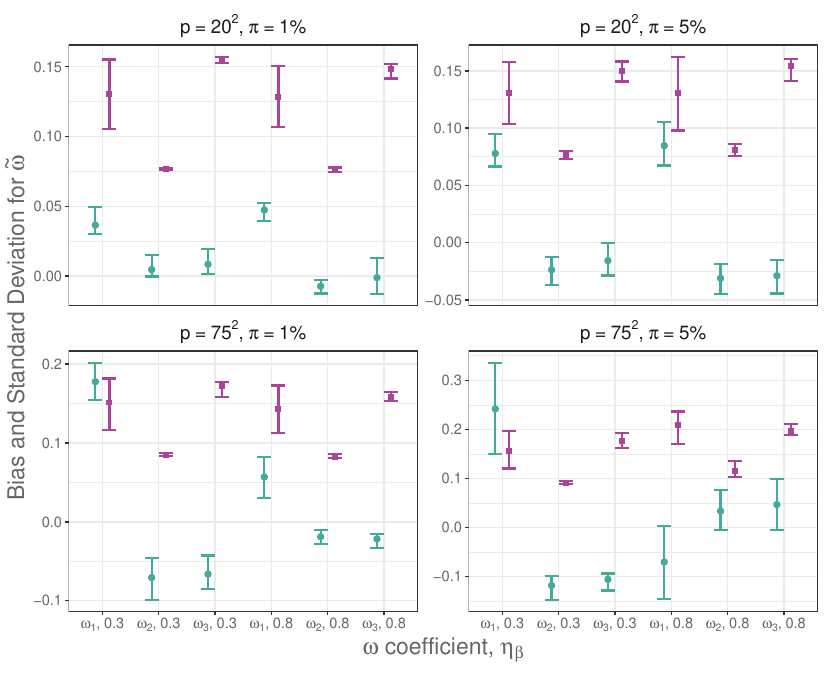} \\
\caption{Bias (green circle) and Standard Deviation (purple squares) of $\tilde{\bomega}$ for H-PROBE. Vertical lines represent the minimum and maximum bias and standard deviations averaged across settings. \label{fig.omega}}
\end{figure}

We also examined the bias and standard deviation of the $\tilde{\bomega}$ coefficient estimates, from the model on the variance. Figure \ref{fig.omega} shows the average bias for the intercept $\omega_1$, the first continuous coefficient $\omega_2$, and the first binary coefficient $\omega_3$, with the vertical bars indicating the minimum and maximum bias across $v$ and $SNR$ settings. The results are displayed by $p$, $\pi$, and $\eta_{\beta}$ settings. In all settings, the bias was lower for first continuous and binary coefficients $\omega_2$ and $\omega_3$, respectively, compared to intercept $\omega_1$. Generally, bias was more pronounced overall in the ultra high-dimensional setting ($p = 75^2$) with more true signals among the available predictors ($\pi=5\%$). The standard deviation for intercept $\omega_1$ and first binary coefficient $\omega_3$ was higher than for continuous coefficient $\omega_2$ in all settings. 

\newpage

\bmsection{Additional Data Analysis Results}\label{sec.sm.data}
\vspace*{12pt}

\begin{figure}[t]
\centering
\includegraphics[width=5.5in]{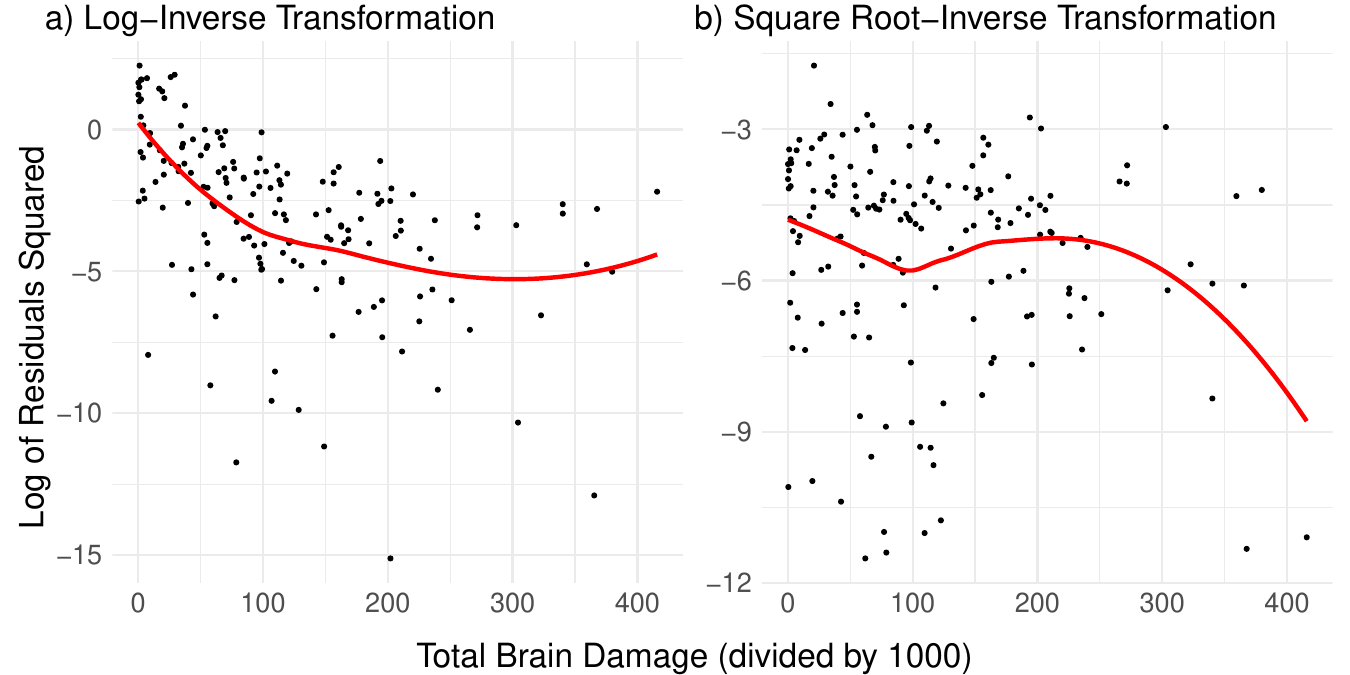} \\
\caption{(a) Log Residuals Squared from PROBE using brain images as predictors and the transformed outcome $AQ_{log-inv}$. (b) Log Residuals Squared from PROBE using brain images as predictors and the transformed outcome $AQ_{sqrt-inv}$\label{fig.resids.aqtransf}}
\end{figure}

This Section provides additional results and information regarding the AQ application and analysis. To ensure that the non-linear relationship between the error variance $\sigma^2_i$ and total brain damage could not be remedied by a transformation of the Aphasia Quotient (AQ) outcome, we performed additional analysis examining transformations of AQ. We examined multiple transformations and provide figures for two of the transformations. We applied log and square root inverse transformations to AQ, $AQ_{log-inv} = log \left( \frac{100-AQ}{100} \right)$ and $AQ_{sqrt-inv} = \sqrt {\left( \frac{100-AQ}{100} \right)}$, and modeled $AQ_{log-inv}, AQ_{sqrt-inv}$ using the PROBE method, a homoscedastic approach. Figure \ref{fig.resids.aqtransf} shows that despite of both transformations, the non-linearity of the relationship between the error variance $\sigma^2_i$ and total brain damage remains. This result remained in other transformations we examined. 

\begin{figure}[t]
\centering
\includegraphics[width=5.5in]{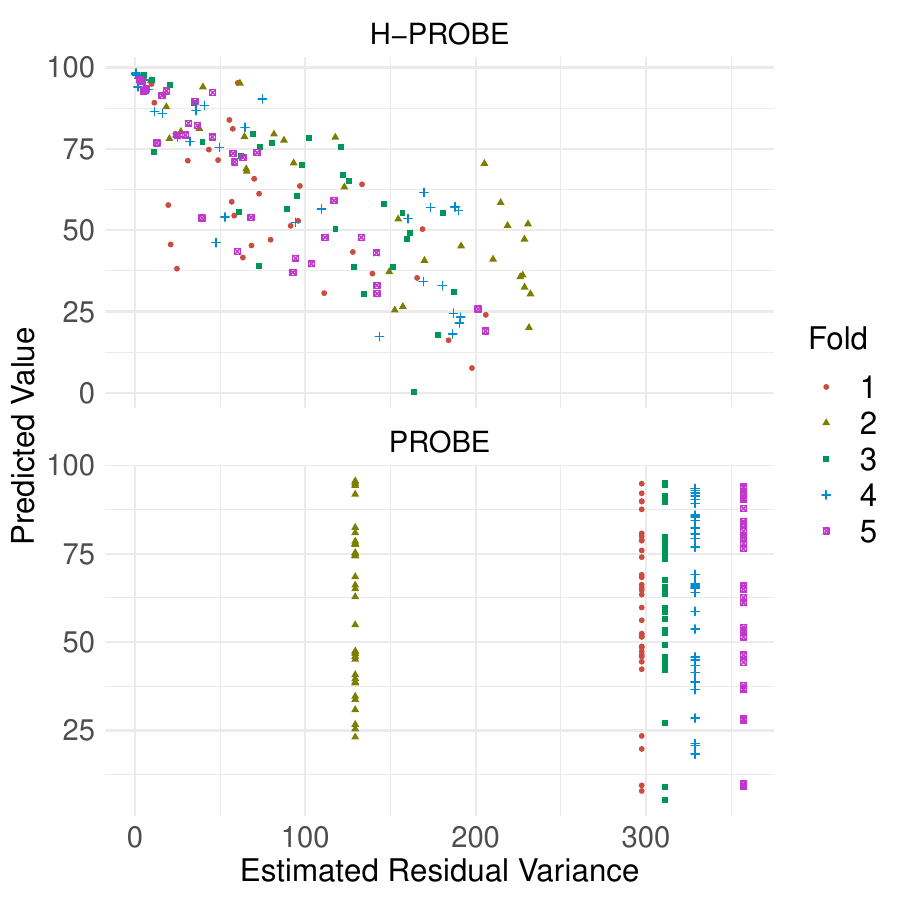} \\
\caption{Predicted Values $\hat{Y}_i$ by estimated $\tilde{\sigma}^2_i$ using H-PROBE and PROBE. Note that within a given fold, $\tilde{\sigma}^2_i$ estimates for observation $i$ are the same when using PROBE. The color legend represents the cross-validation fold in which predictions were obtained. \label{fig.sigma.preds}}
\end{figure}

\begin{figure}[t]
\centering
\includegraphics[width=5.5in]{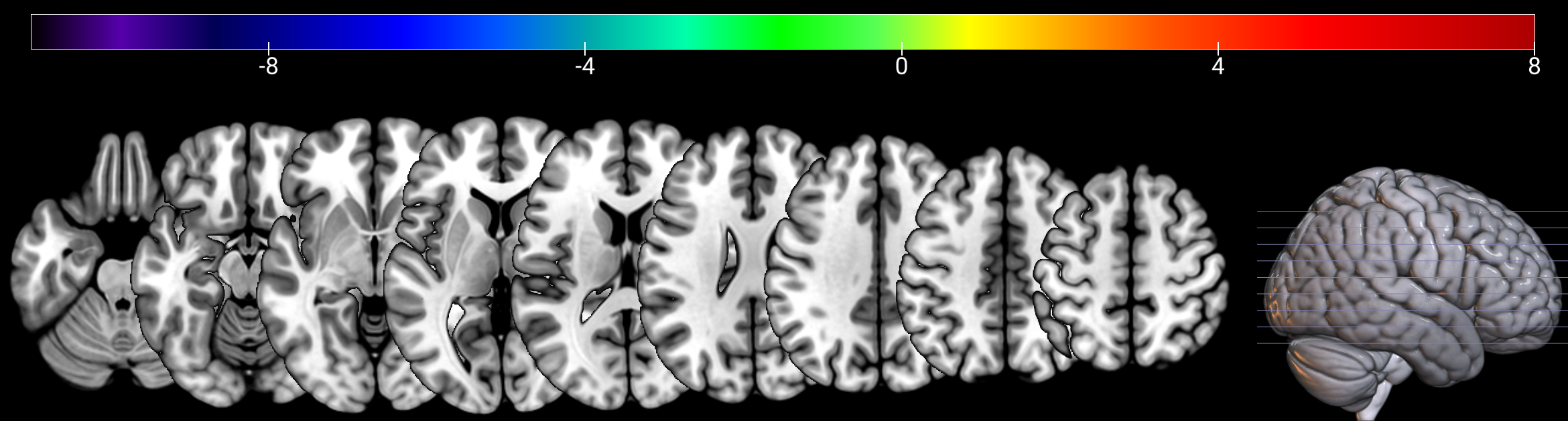} \\
\caption{Brain maps showing position, direction, and magnitude of voxel-specific $\be$ coefficients across different slides for the LASSO model. The color legend represents the magnitude of $\be$ estimates. \label{fig.brainmap.sm}}
\end{figure}

Figure \ref{fig.sigma.preds} shows that H-PROBE provides predictions $\hat{Y}_i$ that have a different range for distinct values of estimated $\tilde{\sigma}^2_i$. This important characteristic of heteroscedastic data is not detected by PROBE. Figure \ref{fig.brainmap.sm} provides per-voxel statistical brain maps for the LASSO method. Since stroke injury is constrained by vasculature, and the presence of a brain injury is an inclusion criteria of the study data, negative $\be$ estimates for a given voxel indicate that the core brain modules related to speech have been spared and are omitted from Figure \ref{fig.brainmap.sm}. The LASSO model overwhelmingly resulted in negative $\be$ estimates, and only one voxel appears on the brain maps, in the fourth brain slide from the left. 

\end{document}